\documentclass{aa}

\usepackage{graphicx}
\graphicspath{{./pictures/}}

\usepackage[varg]{txfonts}

\begin{document}

   \title{Formation of X-ray emitting stationary shocks in magnetized 
   protostellar jets}

   \author{S. Ustamujic
          \inst{1}
          \and
          S. Orlando
          \inst{2}
          \and
          R. Bonito
          \inst{3,2}
          \and
          M. Miceli
          \inst{3,2}
          \and
          A. I. G\'omez de Castro
          \inst{1}
          \and
          J. L\'opez-Santiago
          \inst{4}        
          }

   \institute{S. D. Astronom\'ia y Geodesia, Facultad de Ciencias
   			  Matem\'aticas, Universidad Complutense de Madrid, 28040 Madrid, 
   			  Spain\\
              \email{sustamuj@ucm.es}
         \and
             INAF-Osservatorio Astronomico di Palermo, Piazza del Parlamento 1,
             90134 Palermo, Italy\\
         \and
             Dipartimento di Fisica e Chimica, Universit\`a di Palermo, Via 
             Archirafi 36, 90123 Palermo, Italy \\
         \and
             Dpto. de Astrof\'isica y CC. de la Atm\'osfera, Facultad de
             F\'isica, Universidad Complutense de Madrid, 28040 Madrid, Spain\\
             }

   \date{Received 14 April 2016; accepted 26 July 2016}

 
  \abstract
   {X-ray observations of protostellar jets show evidence of strong shocks 
   heating the plasma up to temperatures of a few million degrees. In some 
   cases, the shocked features appear to be stationary. They are interpreted 
   as shock diamonds.}
   {We aim at investigating the physics that guides the formation of X-ray 
   emitting stationary shocks in protostellar jets, the role of the magnetic 
   field in determining the location, stability, and detectability in X-rays 
   of these shocks, and the physical properties of the shocked plasma.}
   {We performed a set of 2.5-dimensional magnetohydrodynamic numerical 
   simulations modelling supersonic jets ramming into a magnetized medium 
   and explored different configurations of the magnetic field. The model 
   takes into account the most relevant physical effects, namely thermal 
   conduction and radiative losses. We compared the model results with 
   observations, via the emission measure and the X-ray luminosity 
   synthesized from the simulations.}
   {Our model explains the formation of X-ray emitting stationary shocks 
   in a natural way. The magnetic field collimates the plasma at the base of 
   the jet and forms there a magnetic nozzle. After an initial transient, 
   the nozzle leads to the formation of a shock diamond at its exit which is 
   stationary over the time covered by the simulations ($\sim 40-60$~yr; 
   comparable with time scales of the observations). 
   The shock generates a point-like X-ray source located close to the base 
   of the jet with luminosity comparable with that inferred from X-ray 
   observations of protostellar jets. For the range of parameters explored, 
   the evolution of the post-shock plasma is dominated by the radiative 
   cooling, whereas the thermal conduction slightly affects the structure 
   of the shock.}
   {}

   \keywords{Magnetohydrodynamics (MHD) --
                ISM: jets and outflows --
                X-rays: ISM
               }

   \maketitle


\section{Introduction}

   The early stages of a star birth are characterized by a variety of mass
   ejection phenomena, including outflows and collimated jets that are
   strongly related with the accretion process developed in the context of 
   the star-disc interaction. In fact, magnetohydrodynamic (MHD) centrifugal 
   models for jet launching \citep{bla82, pud83} indicated that protostellar 
   jets could provide a valid solution to the angular momentum problem (see 
   \citealt{bac02} for a wider description) via vertical transport along the 
   ordered component of the strong magnetic field threading the disk. 
   According to the widely accepted magneto-centrifugal launching scenario 
   \citep{fer06,gom93}, outflows in young stars are driven from the inner 
   portion of accretion discs and dense plasma from the disc is collimated 
   into jets. This scenario was challenged by several models based on 
   different lines of evidence from the observations suggesting the presence 
   of dust in jets due to material from the outer part of the disc 
   \citep{pod09}. Thus several MHD ejection sites probably coexist in young 
   stars, and the difficulty is to determine the relative contribution of 
   each to the observed jet (for a complete review confronting observations 
   and theory see \citealt{fra14}).
   
   The general consensus is that magnetic fields play a fundamental role in 
   launching, collimating and stabilizing the plasma of jets. This idea was 
   recently corroborated by scaled laboratory experiments that are 
   representative of young stellar object outflows \citep{alb14}. These 
   experiments revealed that stable and narrow collimation of the entire flow 
   can result from the presence of a poloidal magnetic field. A key 
   observational diagnostic to discriminate between different theories could 
   be the detection of possible signatures of rotation in protostellar jets. 
   Pushing the limits of observational resolution, several authors described 
   the detection of asymmetric Doppler shifts in emission lines from opposite 
   borders of the flow in different objects 
   \citep{bac02,woi05,cof04,cof07,cof15}.
   However, this possibility is still undergoing active debate. For 
   example, some inconsistencies in rotation signatures were found at 
   different positions along the jet in a UV study of RW Aur \citep{cof12} 
   and in a near-infrared study of DG Tau \citep{whi14} where systematic 
   transverse velocity gradients could not be identified. Alternative 
   interpretations include asymmetric shocking and/or jet precession 
   \citep[e.g.,][]{sok05,cer06,cor09}.

   Usually jets from young stars are revealed by the presence of a chain of 
   knots, forming the so-called Herbig-Haro (HH) objects. The observations 
   of multiple HH objects showed a knotty structure along the jet axis 
   interpreted as the consequence of the pulsing nature of the ejection of 
   material by the star (e.g. \citealt{rag90,rag07}; \citealt{bon10b,bon10a}; 
   and references therein). After been ejected, the trains of blobs forming 
   the jet move through the ambient medium and they may interact and produce 
   shocks and complex structures that are observed at different wavelengths. 
   In particular, observations showed evidence of faint X-ray emitting 
   sources forming within the jet (e.g.
   \citealt{pra01,fav02,bal03,pra04,tsu04,gud05,ste09}). The origin of this 
   X-ray emission was investigated through hydrodynamic models which have 
   shown that the observations are consistent with the production of strong 
   shocks that heat the plasma up to temperatures of a few million degrees 
   \citep{bon07,bon10b,bon10a}.

   In some cases, the shocked features appeared to be stationary and located 
   close to the base of the jet (e.g. HH154, \citealt{fav06}; DG Tau, 
   \citealt{gud05}). In one of the best-studied X-ray jet, HH154, the X-ray 
   emission consisted of a bright stationary component and a fainter and 
   slightly variable elongated component \citep[see][]{fav06}. On the base 
   of hydrodynamic modelling, this source was interpreted as a shock diamond 
   formed just after a nozzle from which the jet originated \citep{bon11}.
   This scenario was recently supported by scaled laboratory experiments 
   showing the formation of a shock at the base of a laboratory jet that may 
   explain the X-ray emission features observed at the base of some 
   protostellar jets \citep{alb14}.

   Previous MHD models of protostellar jets were aimed at studying the 
   effect of the magnetic field on the dynamical evolution of HH objects 
   (e.g. \citealt{cer97,sul00,sto00}). These studies therefore have focused 
   mainly on the dynamical aspects and the evolution of the jet rather than 
   on obtaining predictions of the emitted X-ray spectrum. Some authors also 
   studied the dynamics considering the pulsing nature of the jets (e.g.
   \citealt{rag98,col06}). These models usually include the radiative cooling 
   but not the effects of thermal conduction, being these studies mostly 
   focussing on the optical emission of jets where the thermal conduction 
   efficiency is lower.

   Here we aim at studying the formation of quasi-stationary X-ray emitting 
   sources close to the base of protostellar jets through detailed 
   2.5-dimensional (2.5D) MHD simulations. We propose a new MHD model which 
   describes the propagation of a jet through a magnetic nozzle which ram 
   with supersonic speed into an initially isothermal and homogeneous 
   magnetized less dense medium. The MHD model takes into account, for the 
   first time, the relevant physical effects, including the radiative losses 
   from optically thin plasma and the magnetic field oriented thermal 
   conduction. The latter is expected to have some effect on the jet 
   dynamics in the presence of plasma at a few MK as that in the presence of 
   strong shocks. In particular the thermal conduction can play an important 
   role in determining the structure of shock diamonds. We compare the 
   results with observations via the emission measure and the X-ray 
   luminosity, synthesized from the simulations. These studies are important 
   to better understand the structure of HH objects and, more specifically, 
   to determine the jet and interstellar magnetic field structure, and may 
   give some insight on the still debated jet ejection and collimation 
   mechanisms.

   The organization of this paper is as follows. In Section 2, we describe 
   the MHD model and the numerical setup. The results of our numerical 
   simulations are described in Section 3. Finally, discussion and 
   conclusions are presented in Section 4.


\section{The model}

   We model the propagation of a continuously driven protostellar jet through
   an initially isothermal and homogeneous magnetized medium. We assume that
   the fluid is fully ionized and that it can be regarded as a perfect gas 
   with a ratio of specific heats $\gamma = 5/3$ (we verified the assumptions 
   used in this paper as described in \citealt{bon07}).
   
   The system is described by the time-dependent MHD equations extended with 
   thermal conduction (including the effects of heat flux saturation) and 
   radiative losses from optically thin plasma. The time-dependent MHD 
   equations written in non-dimensional conservative form are:
   \begin{equation}
      \frac{\partial \rho}{\partial t} 
      + \nabla \cdot (\rho \boldsymbol{u}) = 0,
   \end{equation}
   \begin{equation}
      \frac{\partial \rho \boldsymbol{u}}{\partial t} + \nabla \cdot 
      (\rho\boldsymbol{u}\boldsymbol{u} - \boldsymbol{B}\boldsymbol{B}) 
      + \nabla P_{\mathrm{t}} = 0,
   \end{equation}
   \begin{equation}
      \frac{\partial \rho E}{\partial t}
      + \nabla \cdot [\boldsymbol{u} (\rho E+P_{\mathrm{t}})
      - \boldsymbol{B}(\boldsymbol{u}\boldsymbol{B})] = 
      -\nabla \cdot \boldsymbol{F}_{\mathrm{c}} 
      - n_{\mathrm{e}} n_{\mathrm{H}} \Lambda (T),
   \label{eq.energy}
   \end{equation}
   \begin{equation}
      \frac{\partial \boldsymbol{B}}{\partial t} + \nabla \cdot
      (\boldsymbol{u}\boldsymbol{B}-\boldsymbol{B}\boldsymbol{u}) = 0,
   \end{equation}
   where
   \begin{equation}
      P_{\mathrm{t}} = P + \frac{B^2}{2}, 
      \qquad
      E = \epsilon + \frac{1}{2} u^2 + \frac{1}{2} \frac{B^2}{\rho},
   \end{equation}
   are the total pressure, and the total gas energy (internal energy 
   $\epsilon$, kinetic energy, and magnetic energy) respectively, $t$ is the 
   time, $\rho = \mu m_{\mathrm{H}} n_{\mathrm{H}}$ is the mass density, 
   $\mu = 1.29$ is the mean atomic mass (assuming solar abundances; 
   \citealt{and89}), $m_{\mathrm{H}}$ is the mass of the hydrogen atom, 
   $n_{\mathrm{H}}$ is the hydrogen number density, $\boldsymbol{u}$ is the 
   gas velocity, $\boldsymbol{B}$ is the magnetic field, 
   $\boldsymbol{F}_{\mathrm{c}}$ is the conductive flux, $T$ is the 
   temperature, and $\Lambda (T)$ represents the optically thin radiative 
   losses per unit emission measure derived with the PINTofALE spectral code 
   \citep{kas00} and with the APED V1.3 atomic line database \citep{smi01}, 
   assuming solar metal abundances as before (as deduced from X-ray 
   observations of CTTSs; \citealt{tel07}). We use the ideal gas law, 
   $P = (\gamma-1)\rho\epsilon$.
   
   For the description of the flux-limited thermal conduction in 
   Eq.~\ref{eq.energy}, we adopted the same procedure for smoothly 
   implementing the transition from the classical (\citealt{spi62}) to the 
   saturated (flux-limited) conduction regime (\citealt{cow77}; 
   \citealt{bal82}) suggested by \cite{dal93}. According to \cite{spi62}, 
   the conductive fluxes along and across the magnetic field lines in the
   classical regime are defined as
   \begin{equation}
   \begin{aligned}
      & [q_{\mathrm{spi}}]_{\parallel} = 
      - \kappa_{\parallel} [\nabla T]_{\parallel} 
      \approx -9.2\times10^{-7} T^{5/2} [\nabla T]_{\parallel}, \\
      & [q_{\mathrm{spi}}]_{\perp} = 
      - \kappa_{\perp} [\nabla T]_{\perp} 
      \approx -5.4\times10^{-16} \frac{n^{2}_{\mathrm{H}}}{T^{1/2}B^2} 
      [\nabla T]_{\perp},
   \end{aligned}
   \end{equation}
   where $[\nabla T]_{\parallel}$ and $[\nabla T]_{\perp}$ are the thermal 
   gradients along and across the magnetic field, and $\kappa_{\parallel}$ 
   and $\kappa_{\perp}$ (in units of erg s$^{-1}$ K$^{-1}$ cm$^{-1}$) are 
   the thermal conduction coefficients along and across the magnetic field, 
   respectively. For temperature gradient scales comparable to the electron 
   mean free path, the heat flux is limited and the conductive fluxes along 
   and across the magnetic field lines are given by \citep{cow77}
   \begin{equation}
   \begin{aligned}
      & [q_{\mathrm{sat}}]_{\parallel} = 
      - \mathrm{sign} \left( [\nabla T]_{\parallel} \right) \, 
      5\phi\rho c^{3}_{\mathrm{s}}, \\
      & [q_{\mathrm{sat}}]_{\perp} = 
      - \mathrm{sign} \left( [\nabla T]_{\perp} \right) \, 
      5\phi\rho c^{3}_{\mathrm{s}},
   \end{aligned}
   \end{equation}
   where $\phi \sim 0.3$ (\citealt{giu84}; \citealt{bor89}, and references
   therein) and $c_{\mathrm{s}}$ is the isothermal sound speed.
   The thermal conduction is highly anisotropic due to the presence of the 
   stellar magnetic field, the conductivity being highly reduced in the 
   direction transverse to the field \citep{spi62}. The thermal flux 
   therefore is locally split into two components, along and across the 
   magnetic field lines, 
   $\boldsymbol{F}_{\mathrm{c}} = F_{\parallel}\boldsymbol{i} + 
   F_{\perp}\boldsymbol{j}$, where 
   \begin{equation} \label{thermal}
   \begin{aligned}
      & F_{\parallel} = \left( \frac{1}{[q_{\mathrm{spi}}]_{\parallel}} + 
      \frac{1}{[q_{\mathrm{sat}}]_{\parallel}} \right)^{-1},\\
      & F_{\perp} = \left( \frac{1}{[q_{\mathrm{spi}}]_{\perp}} + 
      \frac{1}{[q_{\mathrm{sat}}]_{\perp}} \right)^{-1},
   \end{aligned}
   \end{equation}
   to allow for a smooth transition between the classical and saturated 
   conduction regime (see \citealt{dal93}; \citealt{orl08,orl10}).

   The calculations were performed using PLUTO \citep{mig07}, a modular, 
   Godunov-type code for astrophysical plasmas. The code provides a 
   multiphysics, multialgorithm modular environment particularly oriented 
   towards the treatment of astrophysical flows in the presence of 
   discontinuities as in our case. The code was designed to make efficient 
   use of massive parallel computers using the message-passing interface 
   (MPI) library for interprocessor communications. The MHD equations are 
   solved using the MHD module available in PLUTO, configured to compute 
   intercell fluxes with the Harten-Lax-Van Leer Discontinuities (HLLD)
   approximate Riemann solver, while second order in time is achieved using 
   a Runge-Kutta scheme. The evolution of the magnetic field is carried out 
   adopting the constrained transport approach \citep{bal99} that maintains 
   the solenoidal condition ($\nabla\cdot B = 0$) at machine accuracy. 
   PLUTO includes optically thin radiative losses in a fractional step 
   formalism \citep{mig07}, which preserves the second time accuracy, as the 
   advection and source steps are at least of the second order accurate; the
   radiative losses ($\Lambda$ values) are computed at the temperature of 
   interest using a table lookup/interpolation method. The thermal conduction 
   is treated separately from advection terms through operator splitting. In 
   particular, we adopted the super-time-stepping technique \citep{ale96} 
   which has been proved to be very effective to speed up explicit 
   time-stepping schemes for parabolic problems when high values of plasma 
   temperature are reached as, for instance, in shocks (see \citealt{orl10} 
   for more details).

\subsection{Numerical setup}
\label{sec:num}

   We adopt a 2.5D cylindrical ($r$, $z$) coordinate system, assuming 
   axisymmetry. We consider the jet axis coincident with the $z$ axis. The 
   computational grid size ranges from $\approx300$ AU to $\approx600$ AU 
   in the $r$ direction and from $\approx900$ AU to $\approx3600$ AU in the 
   $z$ direction, depending on the model parameters. We follow the evolution 
   of the system for at least 40-60 years. These dimensions and time are 
   comparable with those of the observations and are chosen so that we are 
   able to follow the jet collimation and the formation of the shock 
   diamond until a stationary situation is achieved.
   
   We consider an initially isothermal and homogeneous magnetized medium. 
   The initial temperature and density of the ambient are fixed to 
   $T_{\mathrm{a}}=100\,$K and $n_{\mathrm{a}}=100\,$cm$^{-3}$ respectively 
   in all the simulations.  We define a jet, injected into the 
   domain at $z=0$, with a mass ejection rate of $\approx 10^{-8}\,M_{\odot}$ 
   yr$^{-1}$ embedded in an initially axial ($z$) magnetic field. Different 
   magnetic field strengths are investigated. The jet temperature at the 
   lower boundary is assumed to be $T_{\mathrm{j}}=10^6\,$K, in order to 
   obtain values ranging from $10^4$ to $10^5\,$K after the jet expansion in 
   the computational domain. The values used for the model are in good 
   agreement with the observations \citep{fri98,fav02}.

   The mesh is uniformly spaced along the two directions, giving a 
   spatial resolution of $0.5$ AU (corresponding to 120 cells across the 
   initial jet diameter). We performed a convergence test to find the 
   spatial resolution needed to model the physics involved and to resolve the 
   X-ray emitting features. The test consisted on considering the setup for a 
   reference case and performing few simulations with increasing spatial 
   resolution. We found that increasing the resolution adopted in our study 
   by a factor of 2 the results change by no more than 1\%. The domain was 
   chosen according to the physical scales of typical jets from young stars. 
   The adopted resolution is higher than that achieved by current instruments 
   used for the observations of jets, as HST in the optical band and 
   \textit{Chandra} in X-rays. For comparison, the \textit{Chandra} 
   resolution corresponds to $\sim 60$ AU at the distance of HH154 in Taurus 
   ($\sim 140$ pc).

   In all the cases the jet velocity at the lower boundary is 
   oriented along the $z$ axis, coincident with the jet axis, and has a 
   radial profile of the form
   \begin{equation}
   V(r) = \dfrac{V_0}{\nu \, \mathrm{cosh}(r/r_\mathrm{j})^{\omega}-(\nu-1)},
   \end{equation}
   where $V_0$ is the on-axis velocity, $\nu$ is the ambient to jet density 
   ratio, $r_\mathrm{j}$ is the initial jet radius and $\omega = 4$ is the 
   steepness parameter for the shear layer, adjusted so as to achieve a 
   smooth transition of the kinetic energy at the interface between the jet 
   and the ambient medium \citep{bon07}.
   The density variation in the radial direction is given by
   \begin{equation}
   \rho(r) = \rho_\mathrm{j} \left( \nu - 
   \dfrac{\nu-1}{\mathrm{cosh}(r/r_\mathrm{j})^{\omega}} \right),
   \end{equation}
   where $\rho_j$ is the jet density \citep{bod94}.      

   Axisymmetric boundary conditions are imposed along the jet axis 
   (at the left boundary for $r = 0$) in all the cases. At the lower 
   boundary (namely for $z = 0$), inflow boundary conditions (according to 
   the jet parameters given in Table~\ref{parameters}) are imposed for 
   $r \leq r_{\mathrm{j}}$ (where $r_{\mathrm{j}}$ is the jet radius at the 
   lower boundary); for $r \geq r_{\mathrm{j}}$ we checked two different 
   conditions in order to evaluate their effects on our results: 
   (A) equatorial symmetric boundary conditions, where variables are 
   symmetrized across the boundary and tangential component of magnetic 
   field flip sign (in such a way the magnetic field is normal to the 
   boundary for $z = 0$); (B) boundary conditions fixed to the ambient 
   values prescribed at the initial condition (see at the beginning of this 
   section). We tested both A and B boundary conditions for several cases 
   with no relevant variations observed in our results; thus we consider both 
   descriptions equally adapted to our problem. Finally, outflow boundary 
   conditions are assumed elsewhere.

\subsection{Parameters}
\label{sec:description}
 
   Our model solutions depend upon a number of physical parameters, such as, 
   for instance, the magnetic field strength, and the jet density, velocity 
   (including a possible rotational velocity $v_{\varphi}$) and radius. 
   In order to reduce the number of free parameters in our exploration of the 
   parameter space, in every case, we define a jet density and velocity and 
   we fix the jet radius to preserve a mass ejection rate of the order of 
   $10^{-8}\,M_{\odot}$ yr$^{-1}$. Typical outflow rates are found to be 
   between $10^{-7}$ and $10^{-9}$ $M_{\odot}$ yr$^{-1}$ for jets from 
   low-mass classical T Tauri stars (CTTS) \citep{cab07,pod11}.
   We calculate the mass loss rate as 
   $\dot M_{\mathrm{j}} = \int \rho_\mathrm{j} v_\mathrm{j}\,\mathrm{d}A$, 
   where $\rho_\mathrm{j}$ and $v_\mathrm{j}$ are the mass density and jet 
   velocity, respectively, and $\mathrm{d}A$ is the cross sectional area of 
   the incoming jet plasma.

   We define a initially uniform magnetic field along the z axis with values 
   between 0.1 and 5 mG, where $B_{\mathrm{z}}=$ 5 mG is our reference value. 
   This value for the magnetic field strength was chosen according to the 
   value estimated by \cite{bon11} at the exit of a magnetic nozzle close to 
   the base of the jet. It is consistent also with that inferred by 
   \cite{bal03}, namely $B=1-4$ mG, in the context of shocks associated 
   with jet collimation, and by \cite{sch11}, who find $B\approx 6$ mG. 
   These values are also in agreement with those expected at the base 
   of the jet close to the driving source, according to \cite{har07}. As a 
   cross-check we also perform one simulation with $B_{\mathrm{z}}=$ 0.1 mG 
   in order to test the necessity of a minimum magnetic field strength in 
   our model to collimate the plasma and form a shock diamond. In some 
   simulations we consider the plasma of the jet characterised by an angular 
   velocity corresponding to maximum linear rotational velocities of 
   $v_{\varphi,\mathrm{max}} = 1-2\cdot 10^{7}$ cm s$^{-1}$ at the lower 
   boundary. In these cases a toroidal magnetic field component arises and 
   the magnetic field lines are twisted obtaining a helical shaped field.

   The particle number density of the jet, $n_{\mathrm{j}}$, ranges between 
   $10^{4}$ and $10^{5}$ cm$^{-3}$ at the lower boundary . When the 
   jet is injected into the domain the plasma expands and then is 
   collimated by the magnetic field. During this process the density 
   decreases, leading to pre-shock densities of the order of $10^{3}-10^{4}$ 
   cm$^{-3}$, consistent with those inferred by \cite{bal03}. For the jet 
   velocity (also defined at the lower boundary) we explore values between 
   $v_{\mathrm{j}} = 300$ and 1000 km s$^{-1}$ (\citealt{fri05} find 
   velocities of $\sim 500$ km s$^{-1}$ to $\sim 600$ km s$^{-1}$).

   We summarize the parameters of the different models explored in 
   Table~\ref{parameters}. We show the most relevant cases, in particular 
   those where X-ray emission is produced. We define M0 as the reference 
   case because it is the one that most closely reproduces the values of jet 
   density and temperature, and luminosity of the X-ray source derived from 
   the observations \citep{fav02,bal03,gud08,sch11,ski11}.

   \begin{table*}
      \caption[]{Summary of the initial physical parameters characterizing 
      the different models: 
      magnetic field strength along the $z$ axis, $B_{\mathrm{z}}$, 
      maximum linear rotational velocity at the lower boundary, 
      $v_{\varphi,\mathrm{max}}$, 
      jet velocity at the lower boundary, $v_{\mathrm{j}}$, 
      jet density at the lower boundary, $n_{\mathrm{j}}$, 
      initial jet radius, $r_{\mathrm{j}}$, 
      and the mass loss rate calculated at the lower boundary, 
      $\dot M_{\mathrm{j}}$.}
      \label{parameters}
      \centering
      \begin{tabular}{p{3cm} cccccc}
      \hline\hline
      Model  &  $B_{\mathrm{z}}$ (mG)  &  
      $v_{\varphi,\mathrm{max}}$ (km s$^{-1}$)  &  
      $v_{\mathrm{j}}$ (km s$^{-1}$)  &  $n_{\mathrm{j}}$ (cm$^{-3}$)  &  
      $r_{\mathrm{j}}$ (AU)  &  $\dot M_{\mathrm{j}}$ ($10^{-8} M_{\odot}$yr$^{-1}$)\\
      \hline
      \multicolumn{7}{c}{Reference case} \\
      \hline
      M0  &  $5$  &  0  &  $500$  &  10$^{4}$  &  30  &  $1.3$ \\
      \hline
      \multicolumn{7}{c}{Varying magnetic field strength} \\
      \hline
      M1  &  $4$  &  0  &  $500$  &  10$^{4}$  &  30  &  $1.3$ \\
      M2  &  $3$  &  0  &  $500$  &  10$^{4}$  &  30  &  $1.3$ \\
      M3  &  $2$  &  0  &  $500$  &  10$^{4}$  &  30  &  $1.3$ \\
      M4  &  $1$  &  0  &  $500$  &  10$^{4}$  &  30  &  $1.3$ \\
      M5  &  $0.1$  &  0  &  $500$  &  10$^{4}$  &  30  &  $1.3$ \\
      \hline
      \multicolumn{7}{c}{Varying jet rotational velocity} \\
      \hline
      M6  &  $5$  &  $75$  &  $500$  &  10$^{4}$  &  30  &  $1.3$ \\
      M7  &  $5$  &  $112$  &  $500$  &  10$^{4}$  &  30  &  $1.3$ \\
      M8  &  $5$  &  $150$  &  $500$  &  10$^{4}$  &  30  &  $1.3$ \\
      \hline
      \multicolumn{7}{c}{Varying jet velocity} \\
      \hline
      M9  &  $5$  &  0  &  $200$  &  10$^{4}$  &  50  &  $1.3$ \\
      M10  &  $5$  &  0  &  $300$  &  10$^{4}$  &  40  &  $1.4$ \\
      M11  &  $5$  &  0  &  $400$  &  10$^{4}$  &  35  &  $1.3$ \\
      M12  &  $5$  &  0  &  $700$  &  10$^{4}$  &  30  &  $1.8$ \\
      M13  &  $5$  &  0  &  $1000$  &  10$^{4}$ &  20  &  $1.4$ \\
      \hline
      \multicolumn{7}{c}{Varying jet density} \\
      \hline
      M14  &  $5$  &  0  &  $500$  &  10$^{5}$ &  10  &  $1.4$\\
      \hline
      \end{tabular}
   \end{table*}


\section{Results}

\subsection{The reference case}
\label{sec:ref}

\begin{figure*}
      \includegraphics*[width=0.5\textwidth]{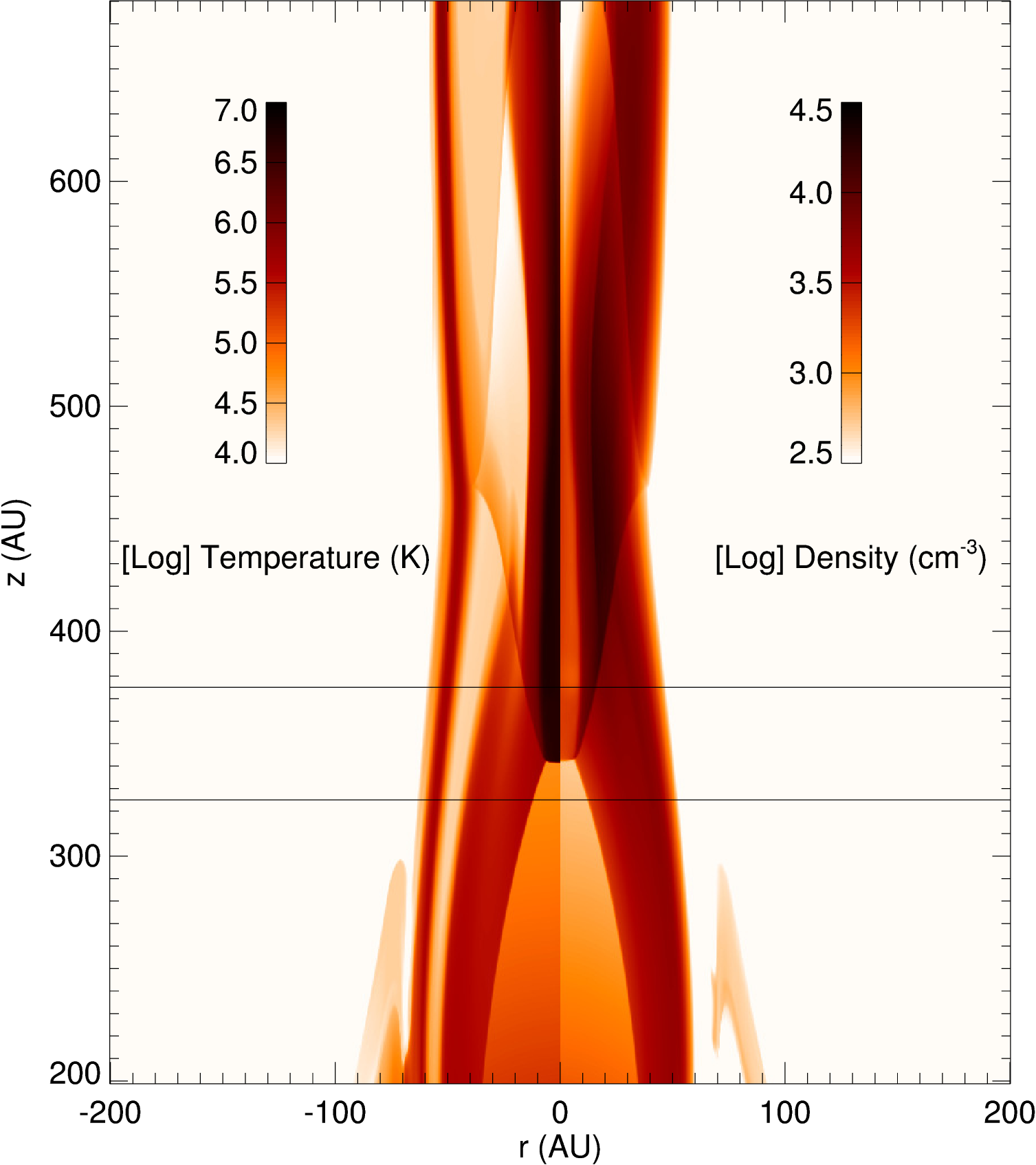}
      \includegraphics*[width=0.5\textwidth]{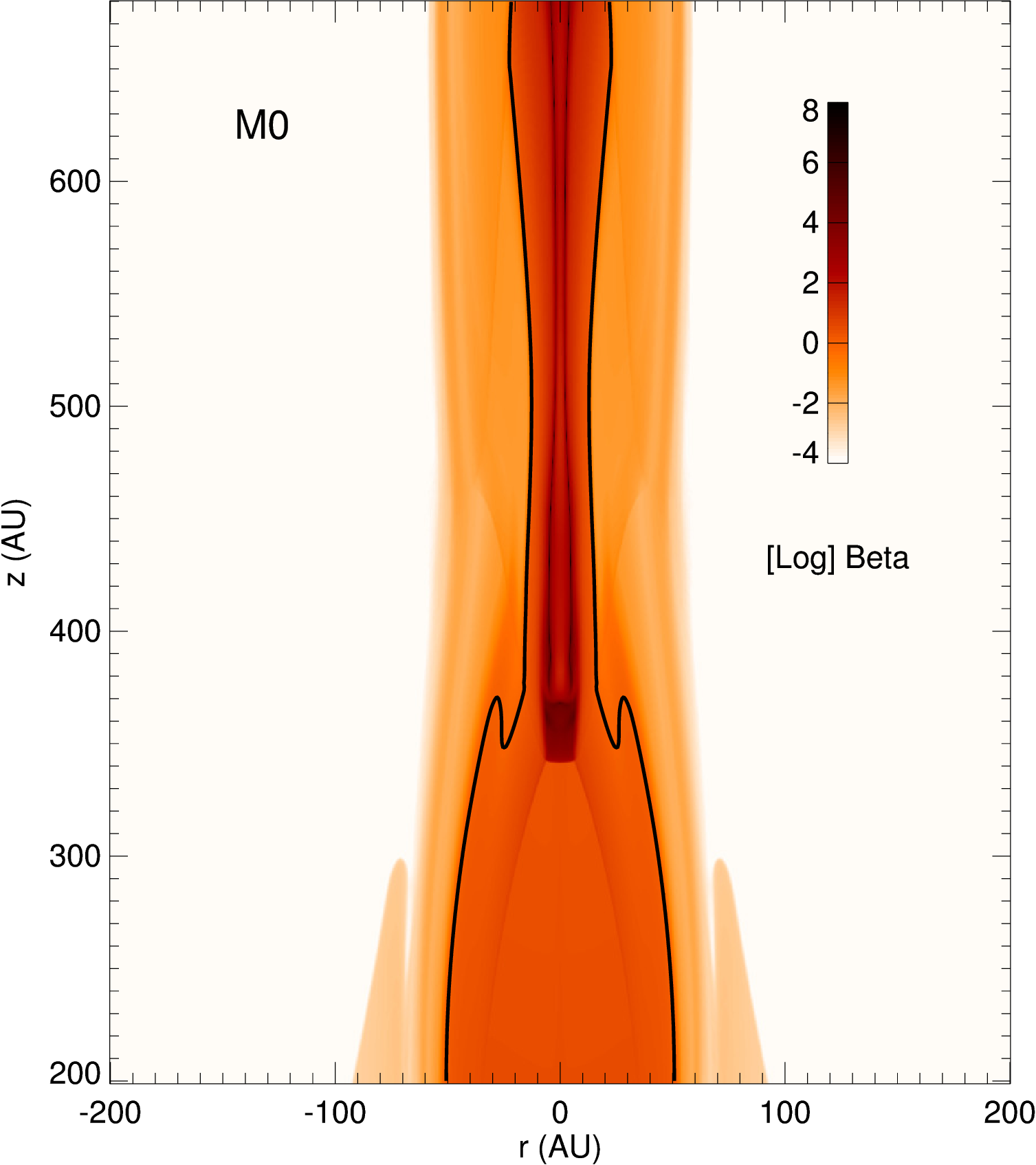}
      \caption{Two-dimensional maps of temperature (left half-panel 
      on the left), density (right half-panel on the left), and plasma 
      $\beta$ (on the right), after $t \approx 50$ years of evolution for 
      the model M0. The contour $\beta=1$ is plotted in black on the right 
      panel. The black horizontal lines plotted on the left panel correspond 
      to $z = 325$ and $z = 375$, which are the radial profiles on the right 
      panels in Figure~\ref{m0_prof}.}
      \label{m0}
   \end{figure*}

   We follow the jet evolution of our reference case M0 (see
   Table~\ref{parameters}) for approximately 50 years to reach a 
   quasi-stationary condition. We show the evolution of the shock diamond 
   in the animation provided online and reporting the 2D spatial 
   distributions of temperature, density and synthesized X-ray emission. 
   The incoming jet, with Mach number 500, initially propagates 
   through the magnetized domain and expands because its dynamic pressure 
   is much larger than the ambient pressure. The jet reaches its maximum 
   expansion at $z \approx 200$ AU (hereafter the collimation point) where 
   its radius is $\approx 90$~AU. During the expansion the jet density and 
   temperature decrease by more than one order of magnitude, while the jet
   velocity remains almost constant. At the same time, the magnetic pressure 
   and field tension increase at the interface between the jet and the 
   surrounding medium, pushing on the jet plasma and forcing it to refocus 
   on the jet axis after the collimation point. As a result the jet is 
   gradually collimated by the ambient (external) magnetic field, reaching a 
   minimum cross-section radius of $\approx 50$~AU at $z \approx 340$~AU 
   (namely $\approx 140$~AU from the collimation point). The flow is 
   compressed by oblique shock waves inclined at an angle to the flow, 
   forming a diamond (see left panel in Fig.~\ref{m0}). A shock wave 
   perpendicular to the jet (the so-called normal shock) forms at 
   $z\approx 340$~AU, when the compressed plasma flows parallel to the jet 
   axis. In this simulation the diamond after $\sim 8$ years of evolution, 
   heats the plasma up to temperatures of a few million degrees, and is 
   stationary until the end of the simulation ($t \approx 50$~yr).

   After the first shock diamond, the flow expands back outward to reduce 
   the pressure again. The flow is expected to be repeatedly compressed and 
   expanded while gradually equalizing the pressure difference between the 
   jet and the ambient, forming a train of shock diamonds along the jet 
   during the process. The final configuration of the magnetic field 
   (after the initial transient) is similar to the solid wall of a 
   conventional nozzle; in particular the magnetic field is very effective 
   in confining the jet plasma, thus acting as a magnetic wall. As an 
   example, in Figure~\ref{m0} (right panel) we show a two-dimensional map
   of the plasma $\beta$ (defined as the ratio of the plasma pressure to the 
   magnetic pressure) for the first shock diamond, located close to the exit 
   of the nozzle. We can distinguish two clearly different regimes: the area 
   close to the jet axis where $\beta>1$, dominated by the jet plasma 
   pressure, in contrast to the rest of the domain where $\beta <1$, 
   dominated by the magnetic field which confines the jet.

\begin{figure}
      \resizebox{\hsize}{!}{\includegraphics*{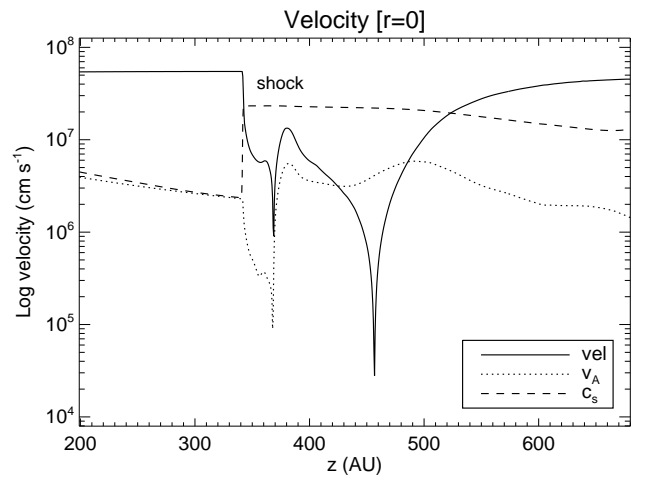}}
      \caption{Magnitude of the jet velocity (solid line), Alfven 
      velocity (dotted line) and sound speed (dashed line) profiles at 
      $r = 0$ and $t \approx 50$ yr for the model M0. The shock is 
      indicated.}
      \label{vel}
   \end{figure}

   In Fig.~\ref{vel} we plot the velocity profiles along the jet axis 
   at the end of the simulation. We compare the magnitude of the jet 
   velocity (solid line), the Alfven velocity (dotted line) and the sound 
   speed (dashed line). In this case, we found that the jet is superalfvenic 
   and supersonic before the shock; immediately after the shock the jet 
   becomes subsonic but not subalfvenic (although it becomes also subalfvenic 
   between $z = 430$ AU and $z = 480$ AU). For $z > 520$ AU and before the 
   next diamond shock, the jet is again superalfvenic and supersonic.

\begin{figure*}
      \resizebox{\hsize}{!}{\includegraphics*{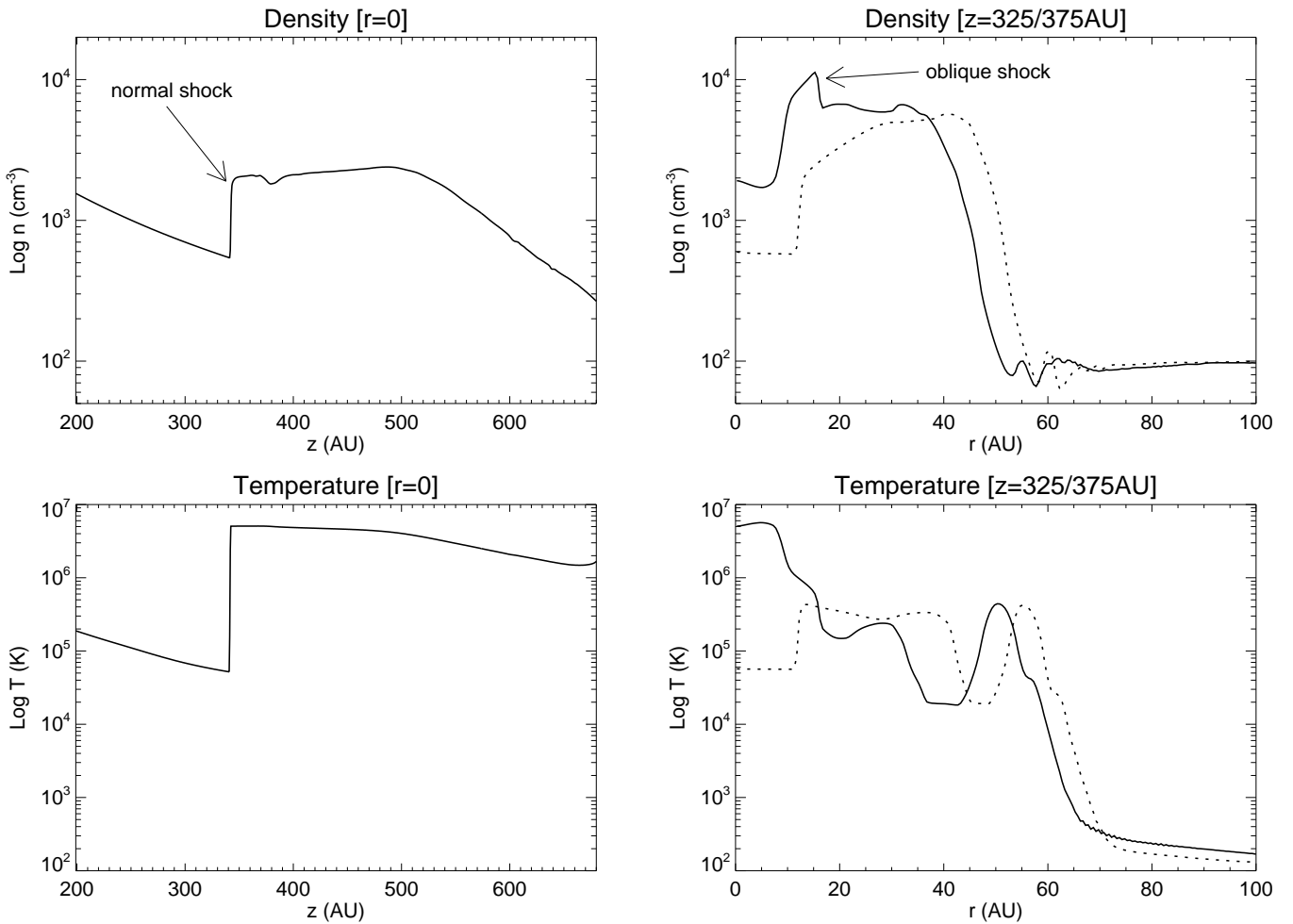}}
      \caption{Density (upper panels) and temperature (lower panels) 
      profiles at $t \approx 50$ yr for the model M0 (see left panel in 
      Fig.~\ref{m0} for the 2D distributions). On the left, $z$ profiles at 
      $r = 0$; on the right, radial profiles at $z = 325$~AU (dotted lines), 
      and at $z = 375$~AU (solid lines). We indicate the position of the 
      normal and oblique shocks in the upper panels.}
      \label{m0_prof}
   \end{figure*}

   We summarize the main physical parameters resulting from the different 
   models in Table~\ref{results}. At the shock diamond, the plasma density 
   and temperature reach maximum values of $\sim 2\cdot10^{4}\,
   \mathrm{cm}^{-3}$ and $\sim 6\cdot10^{6}\,\mathrm{K}$
   respectively in the reference case.  In order to describe the shock X-ray 
   emission contribution, we also calculate the shock temperature (calculated 
   as the density-weighted average temperature) and the shock mass, 
   considering only the cells with $T\geq 10^6$. We obtain a shock 
   temperature of $T_{\mathrm{s}}\approx 2\cdot 10^6$~K and a total mass of 
   $m_{\mathrm{s}}\approx 2\cdot 10^{-9} M_{\odot}$. In 
   Fig.~\ref{m0_prof} we plot the density and temperature profiles along the 
   jet axis and radial profiles few astronomical units before and after the 
   shock (see the corresponding positions marked in Fig.~\ref{m0}); on the 
   left $z$ profiles at $r = 0$ and on the right, radial profiles at 
   $z = 325$ AU (pre-shock) and at $z = 375$ AU (post-shock). The values 
   along the jet axis increase at the normal shock from $\sim 600$ cm$^{-3}$ 
   (pre-shock) to $\sim 2000$ cm$^{-3}$ (post-shock) for the density and from 
   $\sim 5\cdot10^{4}\, \mathrm{K}$ (pre-shock) to 
   $\sim 5\cdot10^{6}\,\mathrm{K}$ (post-shock) for the temperature (see
   left panels in Fig.~\ref{m0_prof}). We note that the material accumulating 
   at the jet border during the collimation determines there an increase of 
   density which is clearly visible in the radial profile of density at 
   $z = 325$~AU, before the shock diamond (see dotted line in the top right 
   panel of Fig.~\ref{m0_prof}). Then the development of the oblique shock is 
   visible in the radial profile of density at $z = 375$~AU, namely after the 
   shock diamond (see solid line in the top right panel of 
   Fig.~\ref{m0_prof}). The temperature decreases as one moves away from the 
   jet axis along the radial direction, with the exception of the jet cocoon, 
   where it increases slightly. We are not expecting any significant X-ray 
   emission from this region as the density is too low. This is confirmed by 
   the results discussed in Sect.~\ref{sec:xray}.

\begin{figure}
      \resizebox{\hsize}{!}{\includegraphics*{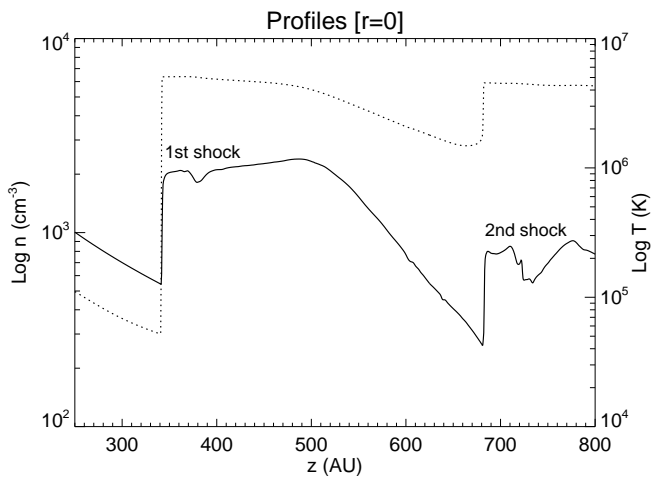}}
      \caption{Density (solid line) and temperature (dotted line) profiles 
      at $r = 0$ and $t \approx 50$ yr for the model M0. Larger scales 
      are used in order to show two consecutive shocks.}
      \label{m0_profiles}
   \end{figure}

   In the following we focus our analysis on the first shock in order to 
   constrain the characteristics of shocks observed at the base of several 
   jets. In fact, we expect that the other shock diamonds predicted by our 
   simulations at larger distances from the base of the jet are fainter in 
   X-rays because of their lower values of density. As an example, in 
   Figure~\ref{m0_profiles} we show the on axis profiles ($r = 0$) of 
   temperature and density for the first two shocks at $z\approx 340$~AU 
   and $z\approx 680$~AU. Before each shock, the density and the temperature 
   decrease due to the jet expansion. The second shock is characterized by a 
   lower density which lead to an even much lower X-ray emission. 
   
\subsection{Exploration of the parameter space}

   We perform a broad exploration of the parameter space defined by the 
   four free parameters: magnetic field strength along the $z$ axis, 
   $B_{\mathrm{z}}$, maximum linear rotational velocity, 
   $v_{\varphi,\mathrm{max}}$, jet velocity, $v_{\mathrm{j}}$, and jet 
   density, $n_{\mathrm{j}}$ (see Sect.~\ref{sec:description} and 
   Table~\ref{parameters}). The aim is to determine the range of parameters 
   leading to the formation of a stationary shock emitting in X-rays. 
   
   We summarize in Table~\ref{results} the main physical parameters of the 
   first shock diamond resulting from the different models defined in 
   Table~\ref{parameters}, namely, the shock position, maximum density and 
   temperature, average temperature, mass and X-ray luminosity. We obtain
   shock temperatures ranging from $\sim 1\cdot 10^{6}$ to 
   $\sim 5\cdot 10^{6}$~K, in excellent agreement with the X-ray 
   observations of \cite{fav02} and \cite{bal03}. The shock luminosity, 
   $L_{\mathrm{X}}$, is calculated considering the whole region including the 
   shock diamond.

   In models M1-M5 we study the role of the magnetic field on the collimation 
   of the jet and the location, stability and detectability in X-rays of the 
   stationary shock. Model M5 considers a very low magnetic field strength 
   (0.1 mG) and, in this case, no shock diamonds form in the computational 
   domain; such a low magnetic field has not enough strength to collimate the 
   jet. In models M3 and M4 the plasma is roughly collimated and a 
   faint shock forms far from the base of the jet. Finally, in models M1 and 
   M2 the shock also forms at large distances from the base of the jet but 
   its X-ray luminosity is higher than that in models M3 and M4 because the 
   shock is more extended in comparison with the reference case and the mass 
   contributing to the X-ray emission is higher (see Table~\ref{results}). 
   In models M6-M8 we study the influence of introducing an angular velocity 
   twisting the magnetic field lines and producing a helicoidal magnetic 
   field. We describe in detail the latter three cases in 
   Sect.~\ref{sec:twist} where we explore the role of the magnetic field 
   twisting. In models M9-M13 we study the effect of changing the jet 
   velocity, $v_{\mathrm{j}}$. In models M9-M11, with lower velocities 
   respect to the reference case, we find that the shock forms closer to the 
   magnetic nozzle than in the reference case and is more extended. Even the 
   shock temperature and density are lower respect to the reference case, 
   they show higher X-ray luminosities as the total mass contributing to the 
   X-ray emission is higher.  However, in models with higher velocities 
   (M12-M13), the shock forms further and has lower luminosity. Lastly, we 
   study the effect of increasing the jet density, $n_{\mathrm{j}}$, in model 
   M14. In this case we obtain a very low luminosity comparing with the other 
   cases, as the mass contributing to the X-ray emission is very low 
   (see Table~\ref{results}).

   Considering the physical parameters from the different models shown in 
   Table~\ref{results}, the most promising cases are M0 (the reference 
   case), M6-M8 (the models considering the magnetic field twisting), and 
   M9-M11 (the models with lower jet velocities). In all these cases a 
   stationary shock forms close to the base of the jet and, as discussed in
   Sect.~\ref{sec:xray}, its X-ray luminosity is comparable with those 
   observed \citep{fav02,bal03,gud08,sch11,ski11}.

   \begin{table*}
      \caption[]{Summary of the main physical parameters resulting from the
      different models: shock position (shock starting position from the 
      beginning of the domain), $d_{\mathrm{s}}$, shock maximum density, 
      $n_{\mathrm{s,max}}$, shock maximum temperature, $T_{\mathrm{s,max}}$, 
      shock temperature (calculated as the density-weighted average 
      temperature), $T_{\mathrm{s}}$, shock mass, $m_{\mathrm{s}}$, and shock 
      X-ray luminosity, $L_{\mathrm{X}}$. For the calculation of the shock 
      temperature and mass, $T_{\mathrm{s}}$ and $m_{\mathrm{s}}$ 
      respectively, we only consider the cells with $T\geq 10^6$ in order to 
      describe the shock X-ray emission contribution.}
      \label{results}
      \centering
      \begin{tabular}{p{1cm} cccccc}
      \hline\hline
      Model  &  $d_\mathrm{s}$ (AU)  &  
      $n_{\mathrm{s,max}}$ ($10^{4}$ cm$^{-3}$)  &  
      $T_{\mathrm{s,max}}$ (MK)  &  $T_{\mathrm{s}}$ (MK)  &  
      $m_{\mathrm{s}}$ ($10^{-9} M_{\odot}$)  &  
      $L_{\mathrm{X}}$ ($10^{28}$ erg s$^{-1}$) \\
      \hline
      M0  &  $340$  &  $1.7$  &  $5.8$  &  $2.1$  &  $1.7$  &  $8.8$ \\
      \hline
      M1  &  $430$  &  $1.7$  &  $4.7$  &  $2.8$  &  $2.4$  &  $9.3$ \\
      M2  &  $570$  &  $1.9$  &  $5.7$  &  $2.5$  &  $2.2$  &  $9.3$ \\
      M3  &  $910$  &  $1.8$  &  $7.0$  &  $2.4$  &  $1.7$  &  $6.3$ \\
      M4  &  $1950$  &  $1.1$  &  $6.3$  &  $2.3$  &  $1.1$  &  $4.7$ \\
      M5  &  \multicolumn{6}{c}{no shock diamond formed} \\
      \hline
      M6  &  $340$  &  $1.7$  &  $5.8$  &  $1.9$  &  $2.0$  &  $11.5$ \\
      M7  &  $340$  &  $1.5$  &  $4.7$  &  $1.9$  &  $2.5$  &  $15.7$ \\
      M8  &  $320$  &  $1.3$  &  $4.7$  &  $2.2$  &  $3.1$  &  $22.1$ \\
      \hline
      M9  &  $160$  &  $0.7$  &  $1.4$  &  $1.2$  &  $2.7$  &  $15.0$ \\
      M10  &  $230$  &  $1.1$  &  $2.3$  &  $1.8$  &  $2.9$  &  $17.8$ \\
      M11  &  $290$  &  $1.4$  &  $3.5$  &  $2.3$  &  $2.1$  &  $11.7$ \\
      M12  &  $530$  &  $2.6$  &  $2.7$  &  $1.5$  &  $2.0$  &  $5.4$ \\
      M13  &  $550$  &  $2.3$  &  $5.2$  &  $1.4$  &  $1.7$  &  $6.2$ \\
      \hline
      M14  &  $450$  &  $302$  &  $3.6$  &  $1.3$  &  $0.06$  &  $ 0.05$ \\
      \hline
      \end{tabular}
   \end{table*}

\subsection{Emission measure distribution vs. temperature}
\label{sec:emvst}

   We derive the distribution of emission measure vs.
   temperature, $EM(T)$, in the temperature range $[10^{4}-10^{8}]$
   K. The $EM(T)$ distribution is an important source of information
   of the plasma components with different temperature contributing
   to the emission and is very useful to compare model results with
   observations. We calculate the $EM(T)$ distribution as follows.
   First we derive the 2D distributions of temperature and density,
   by integrating the MHD equations in the whole spatial domain.
   Then we reconstruct the 3D spatial distribution of these physical
   quantities by rotating the 2D slabs around the symmetry axis
   $z$. For each cell of the 3D domain, we derive the emission
   measure defined as $EM = \int n_{\mathrm{e}} n_{\mathrm{H}}
   \mathrm{d}V$ (where $n_{\mathrm{e}}$ and $n_{\mathrm{H}}$ are
   the electron and hydrogen densities, respectively, and $V$ is
   the volume of emitting plasma).  From the 3D spatial distributions
   of $T$ and $EM$, we derive the $EM(T)$ distribution for the
   computational domain as a whole or for part of it: we consider
   the temperature range $[10^{4}-10^{8}]$ K divided into 80 bins
   equispaced in log $T$; the total $EM$ in each temperature bin
   is obtained by summing the emission measure of all the fluid
   elements corresponding to the same temperature bin.

   \begin{figure}
      \resizebox{\hsize}{!}{\includegraphics*{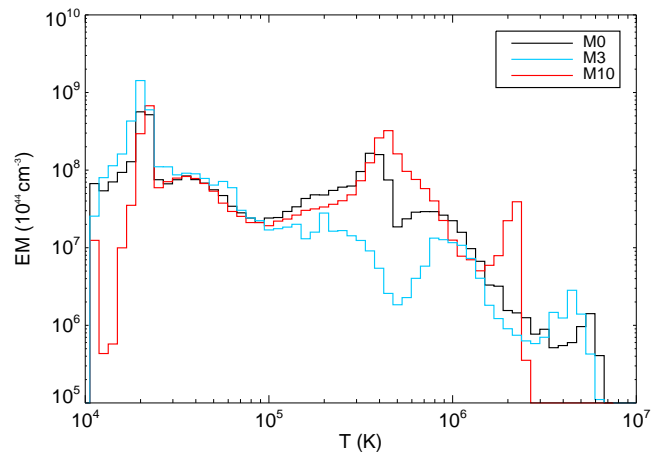}}
      \caption{Emission measure, $EM$, distribution as a function of the 
      temperature, $T$, after $\approx 50$ years of evolution. We compare 
      different models: M0 (the reference case) in black, M3 (one of the 
      less luminous cases) in blue, and M10 (the most luminous case without 
      twisting) in red.}
      \label{m0_emtemp}
   \end{figure}

   Figure~\ref{m0_emtemp} shows the $EM(T)$ for the reference case, M0 (see 
   Sect.~\ref{sec:ref}), after $\approx 50$ years of evolution for the region 
   shown in Fig.~\ref{m0}, corresponding to the first shock. We find that the 
   shape of the $EM(T)$ is characterized by three bumps at 
   $T \approx 2\cdot 10^{4}$~K, $\approx 5\cdot 10^{5}$~K, and 
   $\approx 6\cdot 10^{6}$~K. The latter corresponds to the shock diamond. 
   The $EM$ decreases rapidly above few millions degrees. The same figure 
   compares the $EM(T)$ distribution derived for M0 with those derived for 
   models M3 (blue histogram) and M10 (red histogram). In the former, after a 
   peak of $EM$ at $T\approx 2\cdot 10^4$~K, the $EM$ decreases gradually 
   with the temperature, showing two bumps at $T\approx 10^6$~K and 
   $T\approx 5\cdot 10^6$~K. The $EM$ is lower than that in model M0 and, as 
   a consequence, the X-ray luminosity is expected to be much lower too. 
   Conversely, model M10 shows two intense bumps centred at 
   $T \approx 7\cdot 10^{5}$ K and $T \approx 3\cdot 10^{6}$ K with high 
   values of $EM$. We expect therefore that this model produces higher values 
   of X-ray luminosity.

\subsection{Spatial distribution of X-ray emission}
\label{sec:xray}

   From the 3D spatial distributions of temperature and density reconstructed 
   from the 2.5D simulations (see Sect.~\ref{sec:emvst}), we synthesize the 
   emission in the [0.3-10] keV band to compare the model results with the 
   observations. In particular we derive 2D X-ray images by integrating the 
   X-ray emission along the line of sight (assumed to be perpendicular to the 
   jet axis).

\begin{figure}
      \resizebox{\hsize}{!}{\includegraphics*{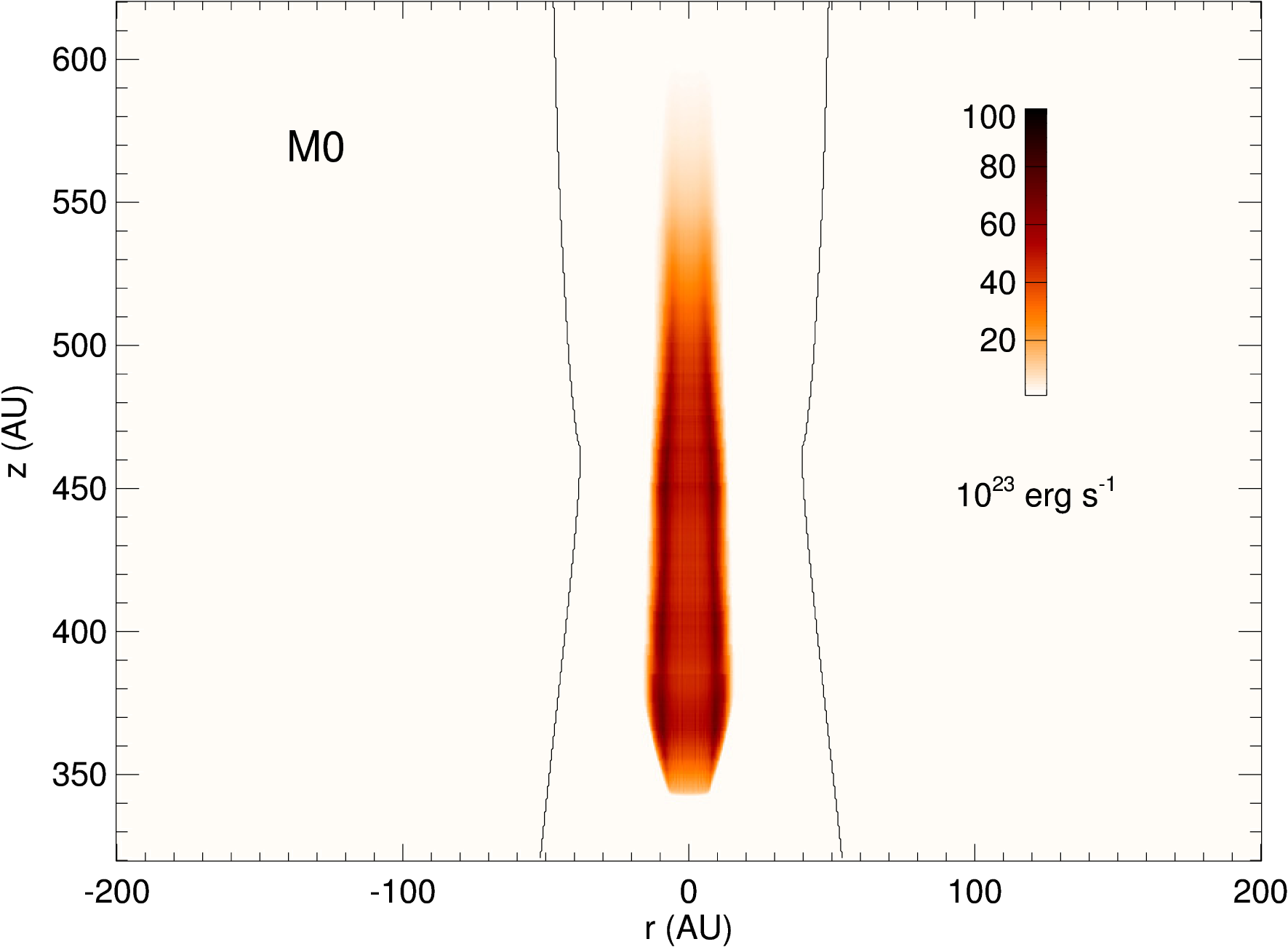}}
      \resizebox{\hsize}{!}{\includegraphics*{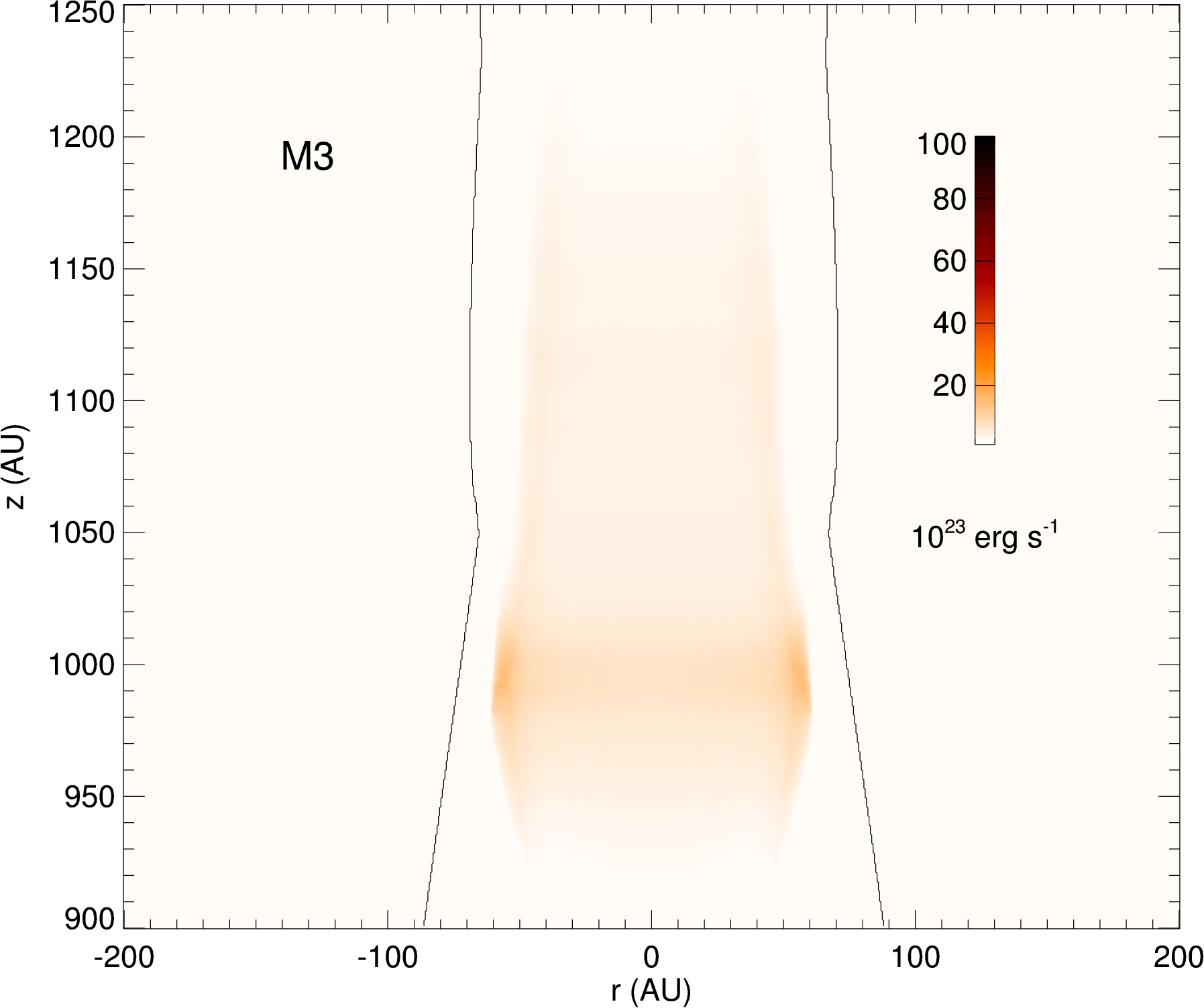}}
      \resizebox{\hsize}{!}{\includegraphics*{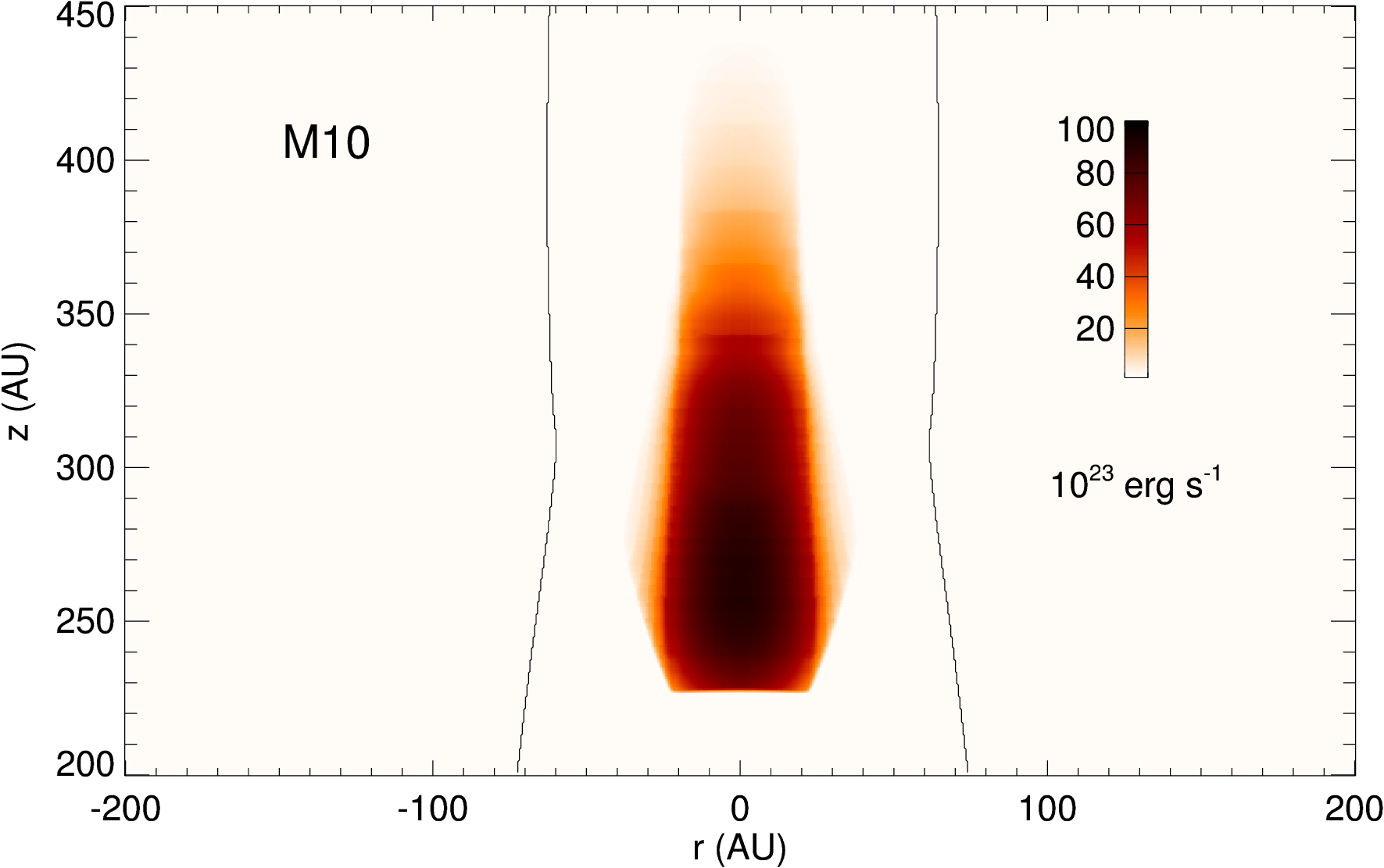}}
      \caption{Spatial distribution of the X-ray emission after $\approx 50$ 
      years of evolution. We compare different models: M0 (the reference 
      case) upper panel, M3 (one of the less luminous cases) middle panel, 
      and M10 (the most luminous case without twisting) in the lower panel. 
      We plot the contour of the jet in black in each case.}
      \label{m0_emlum}
   \end{figure}

   Figure~\ref{m0_emlum} shows the spatial distribution of the X-ray emission 
   for the reference case, M0 (see Sect.~\ref{sec:ref}), for M3 (one of the 
   less luminous cases) and M10 (the most luminous case without twisting), 
   after 48 years of evolution. The X-ray source appears as a single 
   elongated source corresponding to the shock diamond region 
   developing as a result of the jet collimation (see discussion in 
   Sect.~3.1). Note the different scale for the $z$ axis in the three panels 
   of the figure and also with respect to Fig.~\ref{m0}. After an initial 
   transient ($\sim 10$ yr), all these sources appear to be stationary over
   the time covered by the simulations ($\sim 50$ yr; see the movie provided 
   online). Almost all the X-ray emission originates from the shock diamond, 
   whereas the contribution arising from the jet envelope is negligible in 
   all the cases.

   The X-ray total shock luminosity, $L_{\mathrm{X}}$, in the [0.3-10] keV 
   band, derived for the model M0 is $\sim 9\cdot 10^{28}$ erg s$^{-1}$ 
   (see Table~\ref{results}). The spatial X-rays distribution for the model 
   M10 has wider shape and higher total shock luminosity, namely 
   $\sim 2\cdot 10^{29}$ erg s$^{-1}$. Finally, the spatial X-rays 
   distribution for the model M3 is fainter and more extended, with a total 
   shock luminosity of $\sim 6\cdot 10^{28}$ erg s$^{-1}$. These values are 
   comparable with those detected in several HH objects
   \citep{fav02,bal03,gud08,sch11,ski11}, ranging between $2.4\cdot 10^{28}$ 
   and $3\cdot 10^{29}$ erg s$^{-1}$.
   
   Note that the X-ray luminosity, $L_{\mathrm{X}}$, is calculated assuming 
   that the plasma is optically thin. In order to verify our assumption, we 
   estimated a hydrogen column density 
   $N_\mathrm{H} \approx 9\cdot 10^{17}$ cm$^{-2}$, assuming a shock 
   thickness of 60 AU and a density of $10^3$ cm$^{-3}$. Thus, the X-ray 
   absorption is negligible.
   
   It is also worth noting that hydrodynamic models predict significant X-ray 
   emission by jets less dense than the ambient medium (light jets), whereas 
   jets denser than the ambient (heavy jets) produce X-ray luminosities even 
   several orders of magnitude lower than those observed 
   (e.g. \citealt{bon11}). Our results show that the magnetic field can 
   enhance the X-ray luminosity from jets, leading to X-ray luminosities 
   comparable with those observed even in the case of heavy jets. Here we 
   suggest that the magnetic field plays a fundamental role in observed X-ray 
   emitting heavy jets through efficient magnetic collimation of the flow in 
   shock diamonds. 

\subsection{Role of the jet rotational velocity}
\label{sec:twist}

\begin{figure}
      \resizebox{\hsize}{!}{\includegraphics*{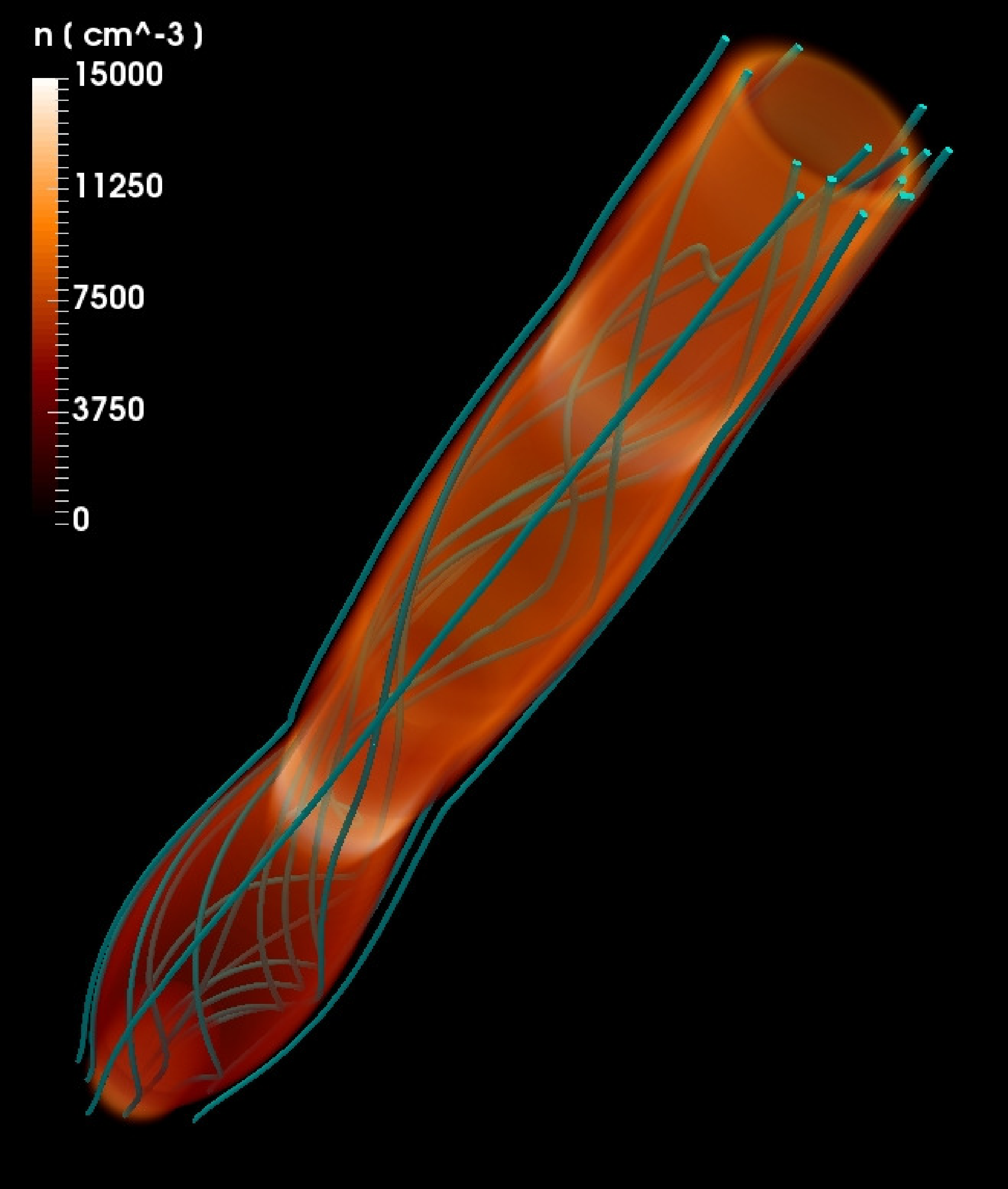}}
      \caption{3D representation of the density distribution and the magnetic 
      field configuration (blue lines) for the model M8 after $\approx 50$ 
      years of evolution.}
      \label{twist}
   \end{figure}

   In models M6, M7 and M8 (see Table~\ref{parameters}) we study the 
   influence of introducing an angular velocity to the jet. We define a 
   constant angular velocity producing maximum linear rotational velocity 
   values, $v_{\varphi,\mathrm{max}}$, ranging from 75 to 150 km s$^{-1}$ 
   at the lower boundary\footnote{In the presence of an angular velocity, 
   we found that the boundary conditions B (see Sect.~\ref{sec:num}) are the 
   most appropriate, because the boundary conditions A can produce numerical 
   artefacts at the base of the jet.}. The jet rotation produces a toroidal 
   component at the initially axial magnetic field achieving a helicoidal
   configuration during the simulation. The final configuration of the 
   magnetic field depends on the rotational velocity of the plasma; the 
   higher the velocity, the more twisted the magnetic field lines. In 
   Figure~\ref{twist} we show a 3D representation of the density distribution 
   and magnetic field configuration for model M8, the one with the highest 
   rotational velocity of the cases explored (see Table~\ref{parameters}). 
   The magnetic field lines are twisted in a helicoidal configuration,
   contributing to the jet collimation.

\begin{figure*}
      \includegraphics*[width=0.33\textwidth]{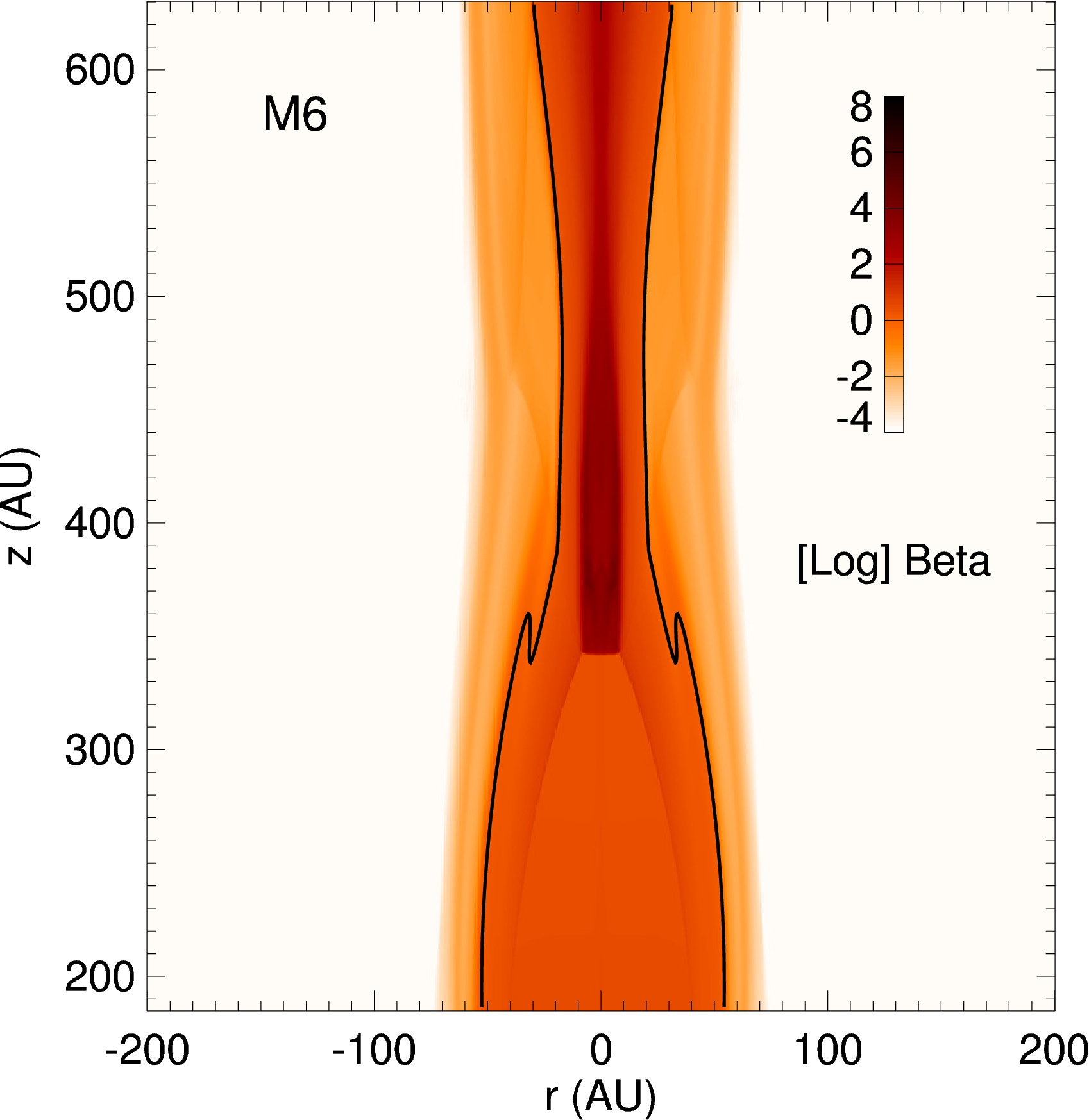}
      \includegraphics*[width=0.33\textwidth]{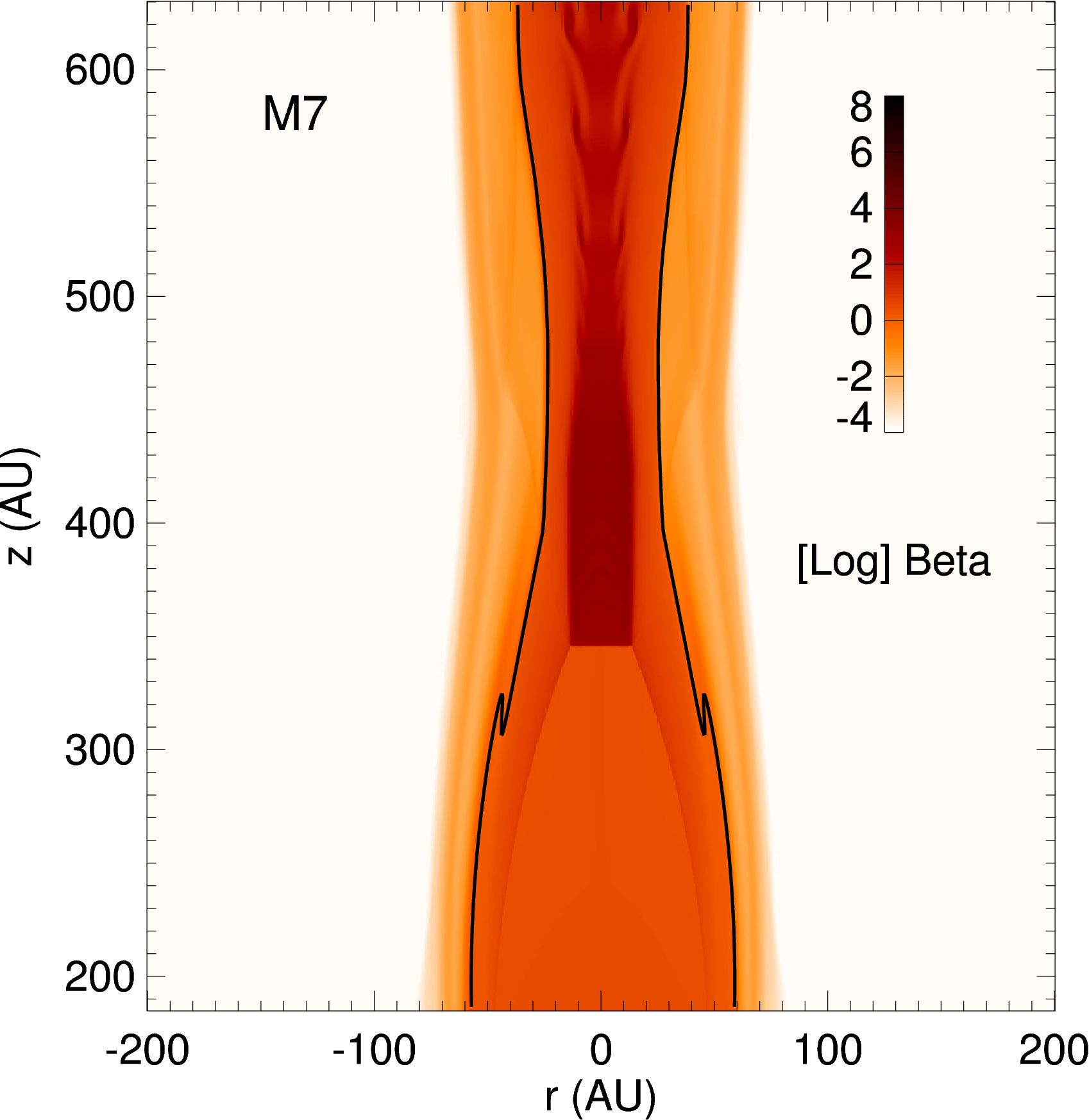}
      \includegraphics*[width=0.33\textwidth]{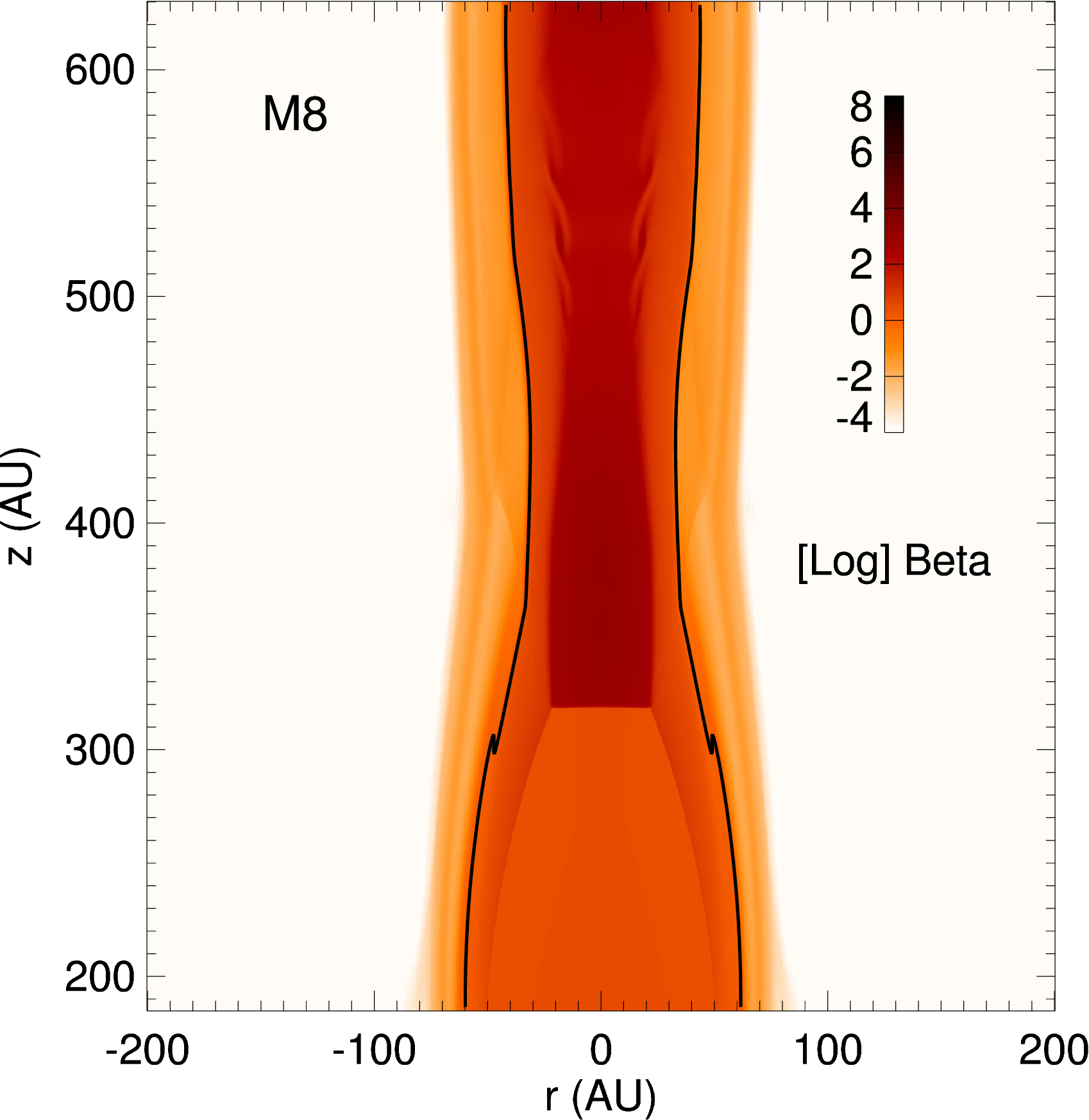}
      \caption{Two-dimensional maps of plasma $\beta$ after $\approx 50$ 
      years of evolution. Different models are compared: M6 (the model with 
      the lowest rotational velocity) left panel, M7 (the model with an 
      intermediate rotational velocity) middle panel, and M8 (the model with 
      the highest rotational velocity) right panel. The contour $\beta=1$ is 
      plotted in black in every case.}
      \label{tw_beta}
   \end{figure*}

   In Figure~\ref{tw_beta} we show two-dimensional maps of the plasma 
   $\beta$ for the three models. We can distinguish two different regimes 
   in all the cases: the area close to the jet axis where $\beta >1$, 
   dominated by the jet plasma pressure, in contrast to the rest of the 
   domain where $\beta <1$, dominated by the magnetic field. According to 
   the maps, the higher the rotational velocity the larger the region with 
   $\beta >1$, dominated by the jet plasma pressure, and the stronger and 
   more extended the shock. This is basically due to the increase of the 
   dynamic pressure in the jet interior.

\begin{figure}
      \resizebox{\hsize}{!}{\includegraphics*{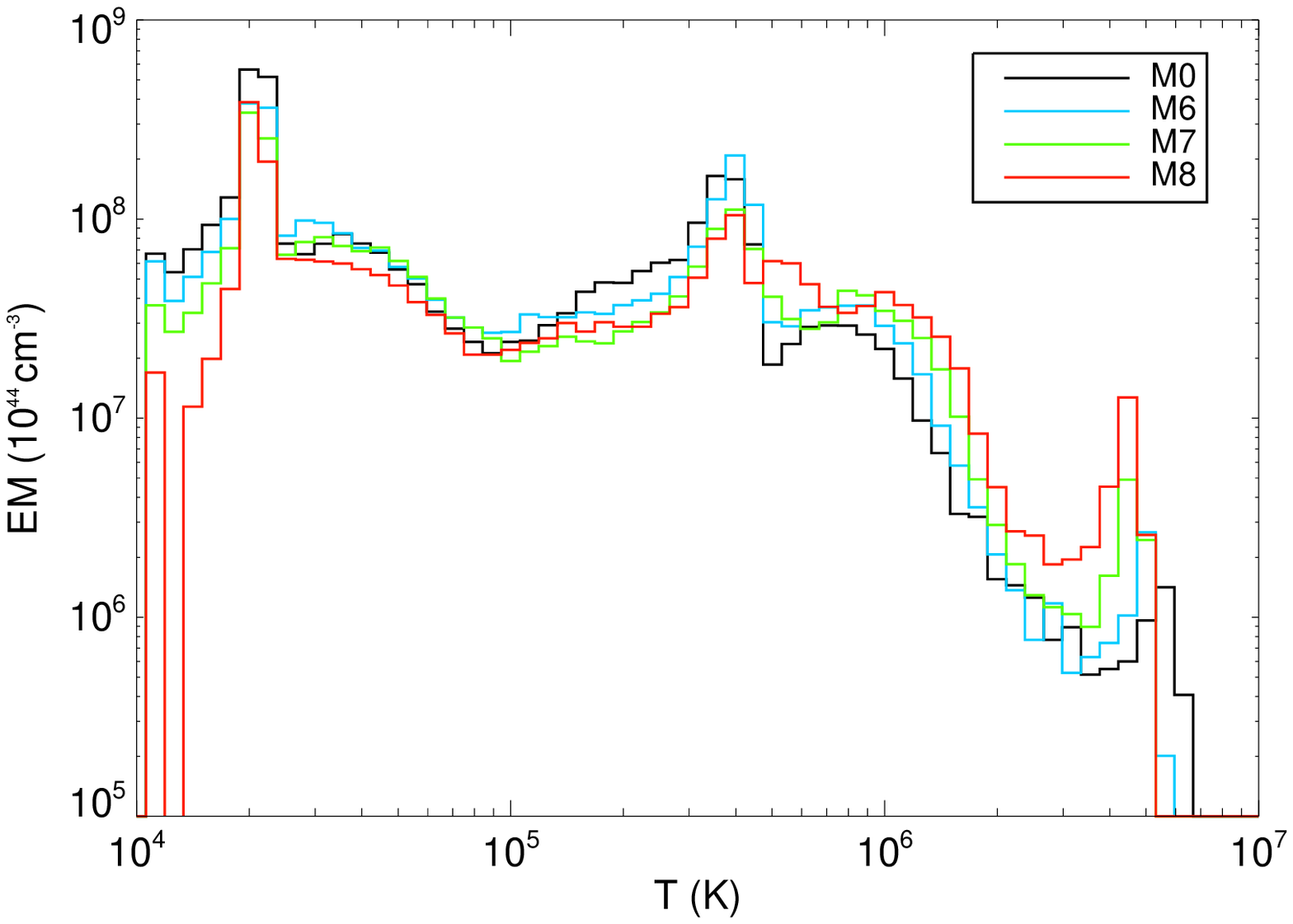}}
      \caption{Emission measure, $EM$, distribution as a function of the 
      temperature, $T$, after $\approx 50$ years of evolution. We compare 
      different models: M0 (the reference case) in black, M6 (the model with 
      the lowest rotational velocity) in blue, M7 (the model with an 
      intermediate rotational velocity) in green, and M8 (the model with the 
      highest rotational velocity) in red.}
      \label{tw_emtemp}
   \end{figure}

   Figure~\ref{tw_emtemp} shows the emission measure distribution vs. 
   temperature, $EM(T)$, calculated as explained in Section~\ref{sec:emvst}. 
   We compare the reference case, M0, with the models M6, M7 and M8 (see 
   Table~\ref{parameters}); this allows us to explore the effect of an 
   increasing angular velocity and the expected emission from the jet. We 
   find that, increasing the rotational velocity $v_{\varphi,\mathrm{max}}$, 
   the $EM$ decreases for temperatures $T<6\cdot 10^5$ K and increases for 
   higher temperatures. Thus we expect that, increasing 
   $v_{\varphi,\mathrm{max}}$, the shock diamond is a brighter X-ray source 
   with higher X-ray luminosity.

\begin{figure}
      \resizebox{\hsize}{!}{\includegraphics*{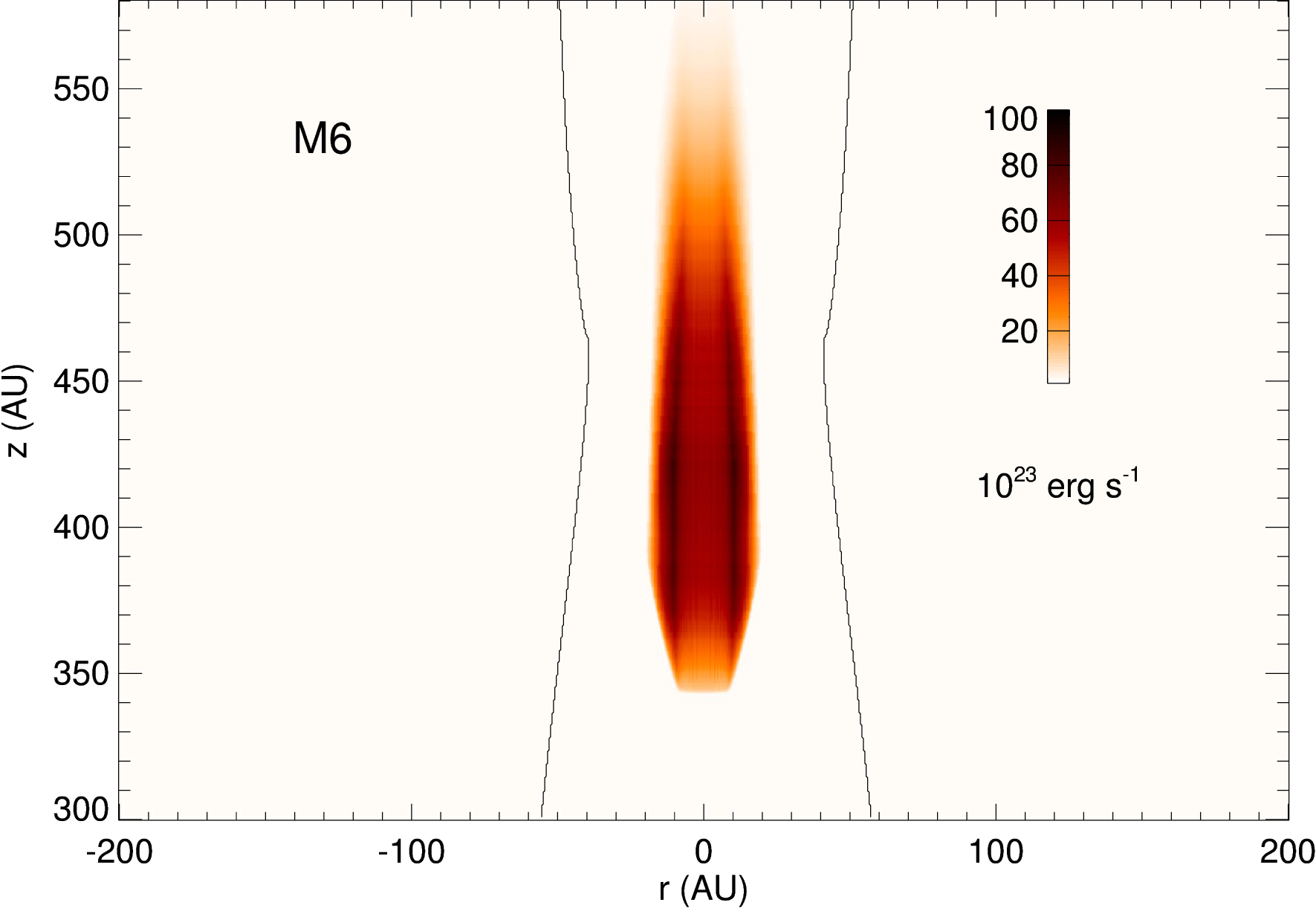}}
      \resizebox{\hsize}{!}{\includegraphics*{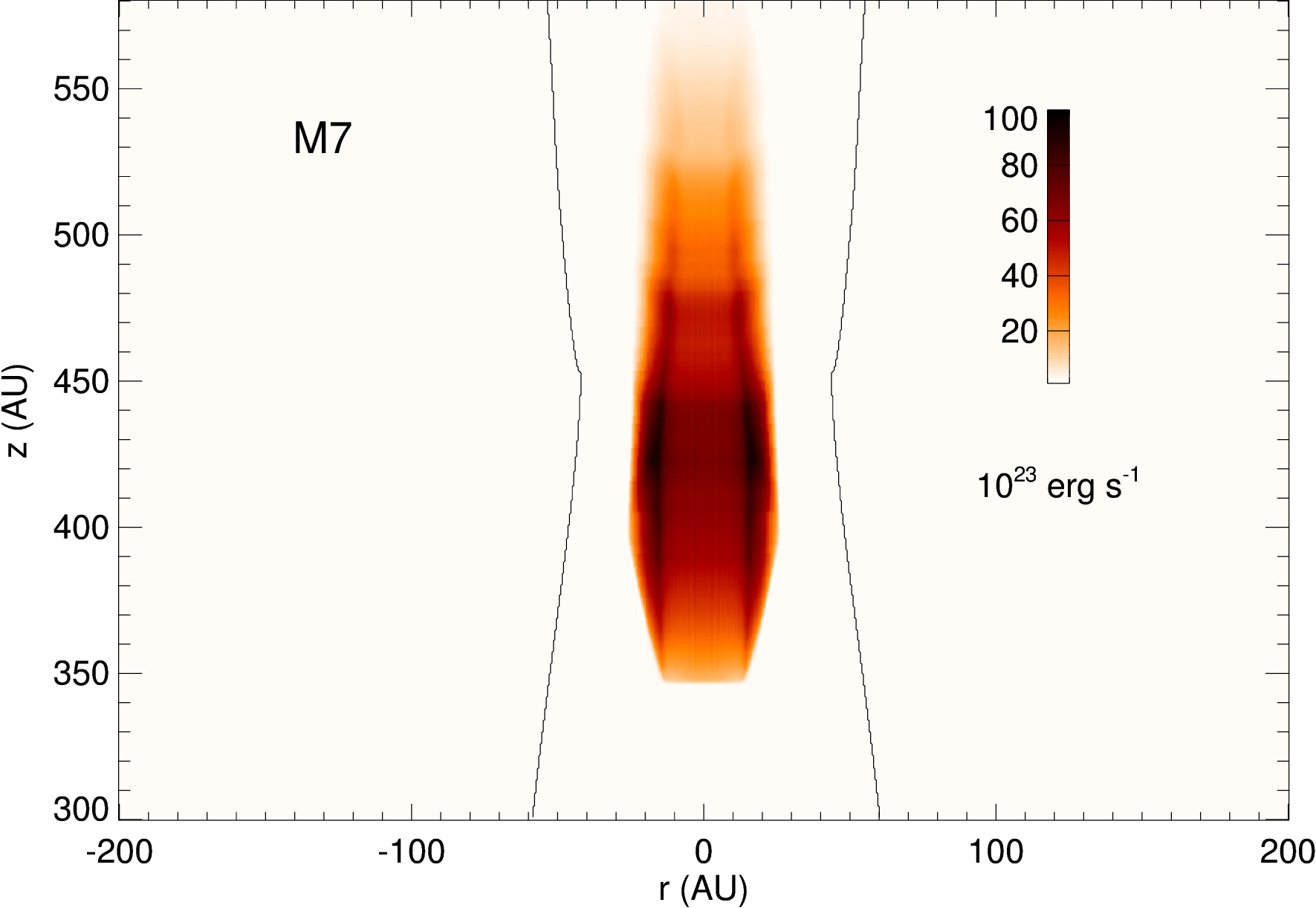}}
      \resizebox{\hsize}{!}{\includegraphics*{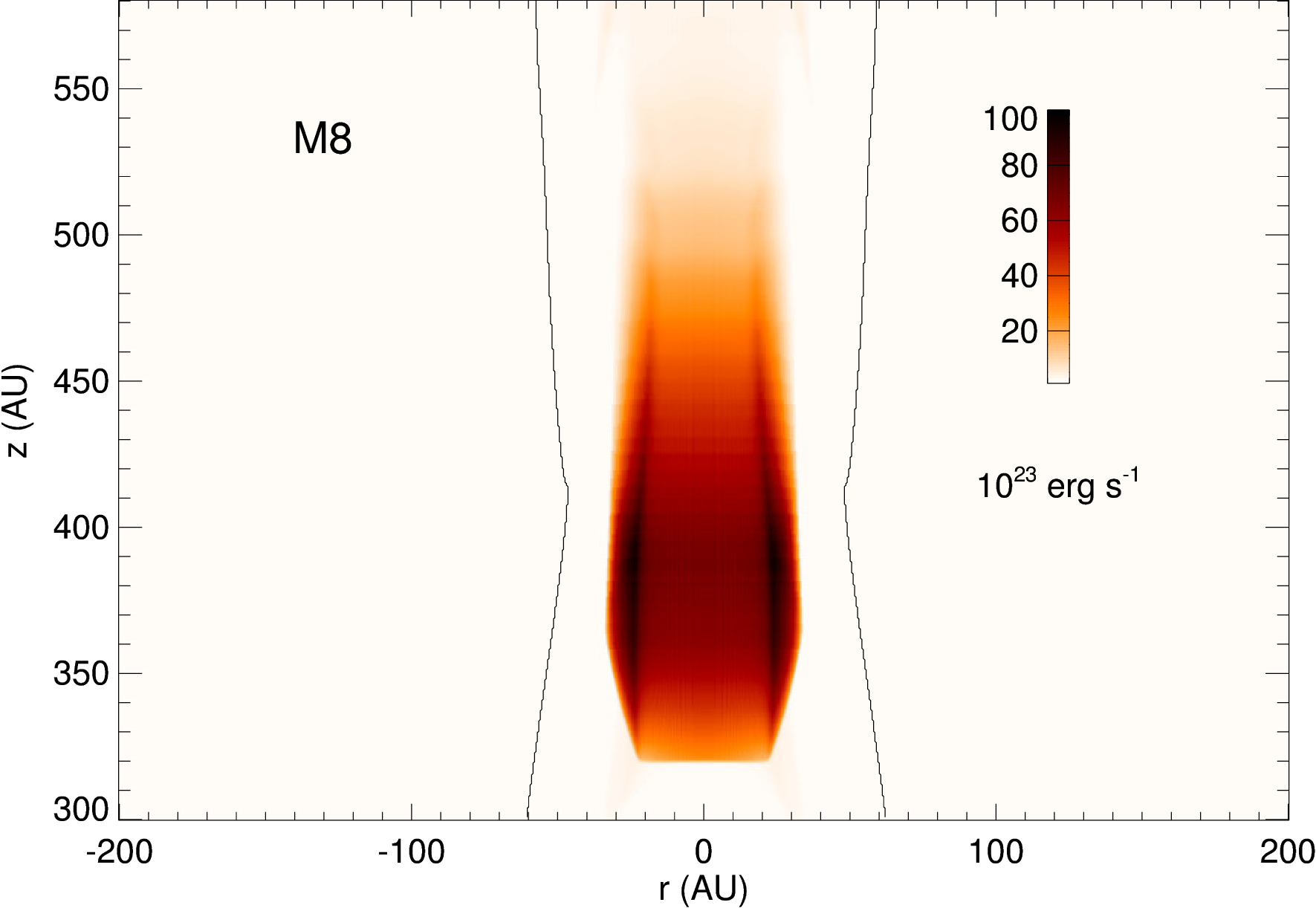}}
      \caption{Spatial distribution of the X-ray emission after $\approx 50$ 
      years of evolution. Different models are compared: M6 (the model with 
      the lowest rotational velocity) upper panel, M7 (the model with an 
      intermediate rotational velocity) middle panel, and M8 (the model with 
      the highest rotational velocity) lower panel. We plot the contour of 
      the jet in black in each case.}
      \label{tw_emlum}
   \end{figure}

   Figure~\ref{tw_emlum} shows the spatial distribution of the X-ray emission 
   for the three different models with twisting (see Table~\ref{parameters}), 
   derived as explained in Section~\ref{sec:xray}. We find that the X-ray 
   emission source, corresponding to the shock diamond, is closer to the base 
   of the jet and more extended for models with higher jet rotation 
   velocities. The X-ray total shock luminosity, $L_{\mathrm{X}}$, in the
   [0.3-10] keV band, is larger for higher rotational velocity of the jet 
   (see Table~\ref{results}), as the total mass contributing to the X-ray 
   emission is higher. These values are consistent with those detected in 
   several HH objects \citep{fav02,bal03,gud08,sch11,ski11}.

\begin{figure*}
      \includegraphics*[width=0.33\textwidth]{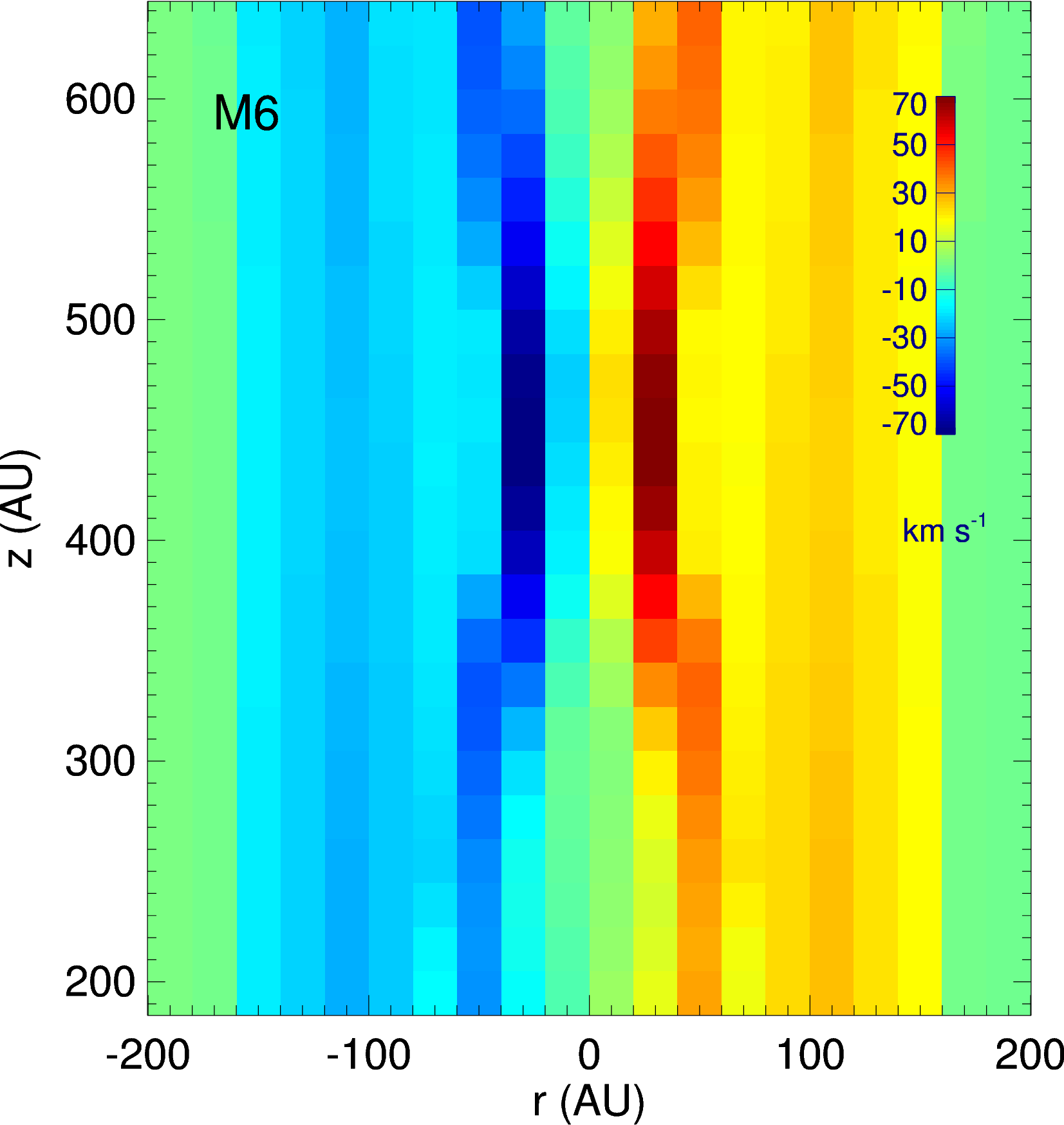}
      \includegraphics*[width=0.33\textwidth]{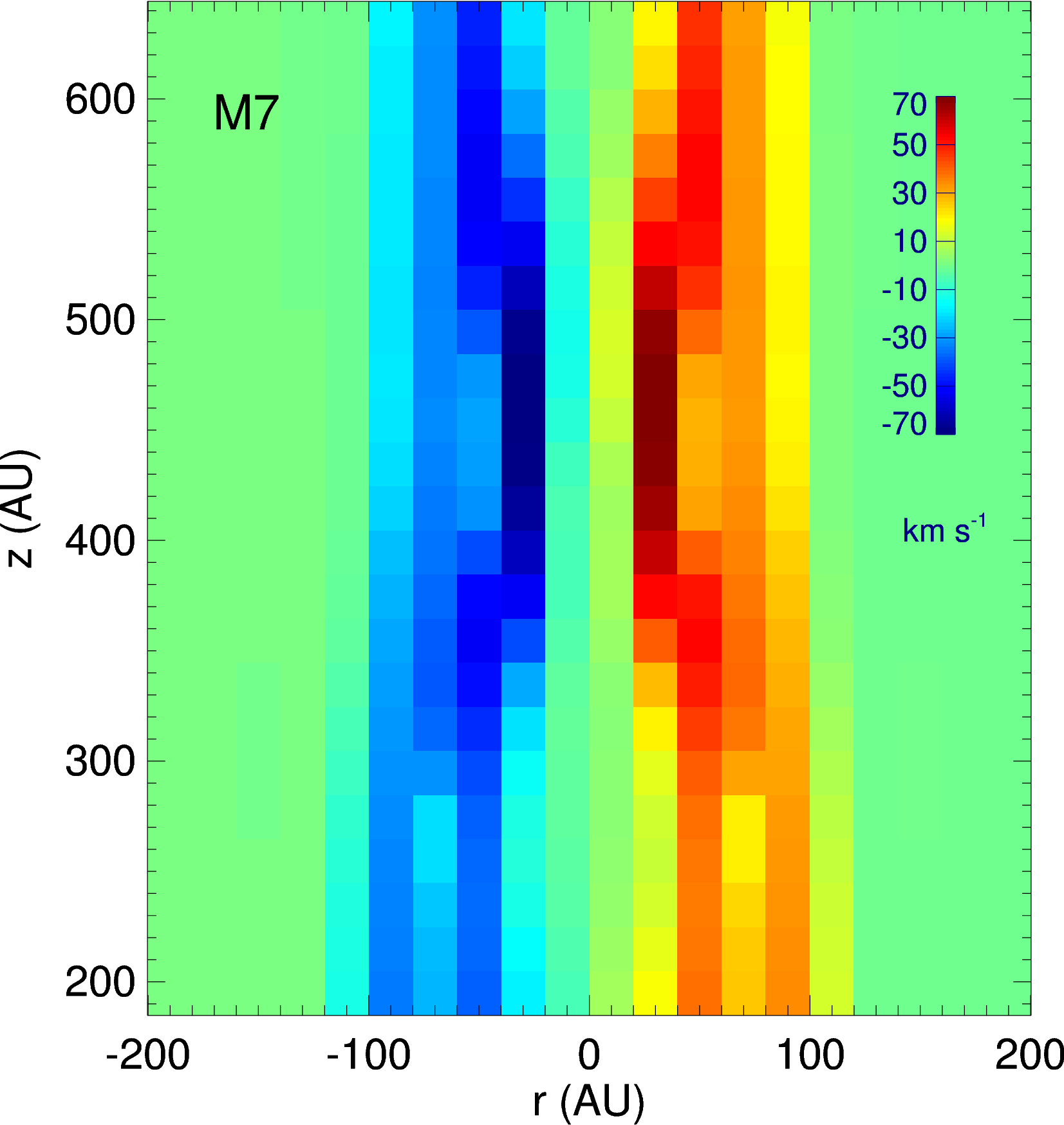}
      \includegraphics*[width=0.33\textwidth]{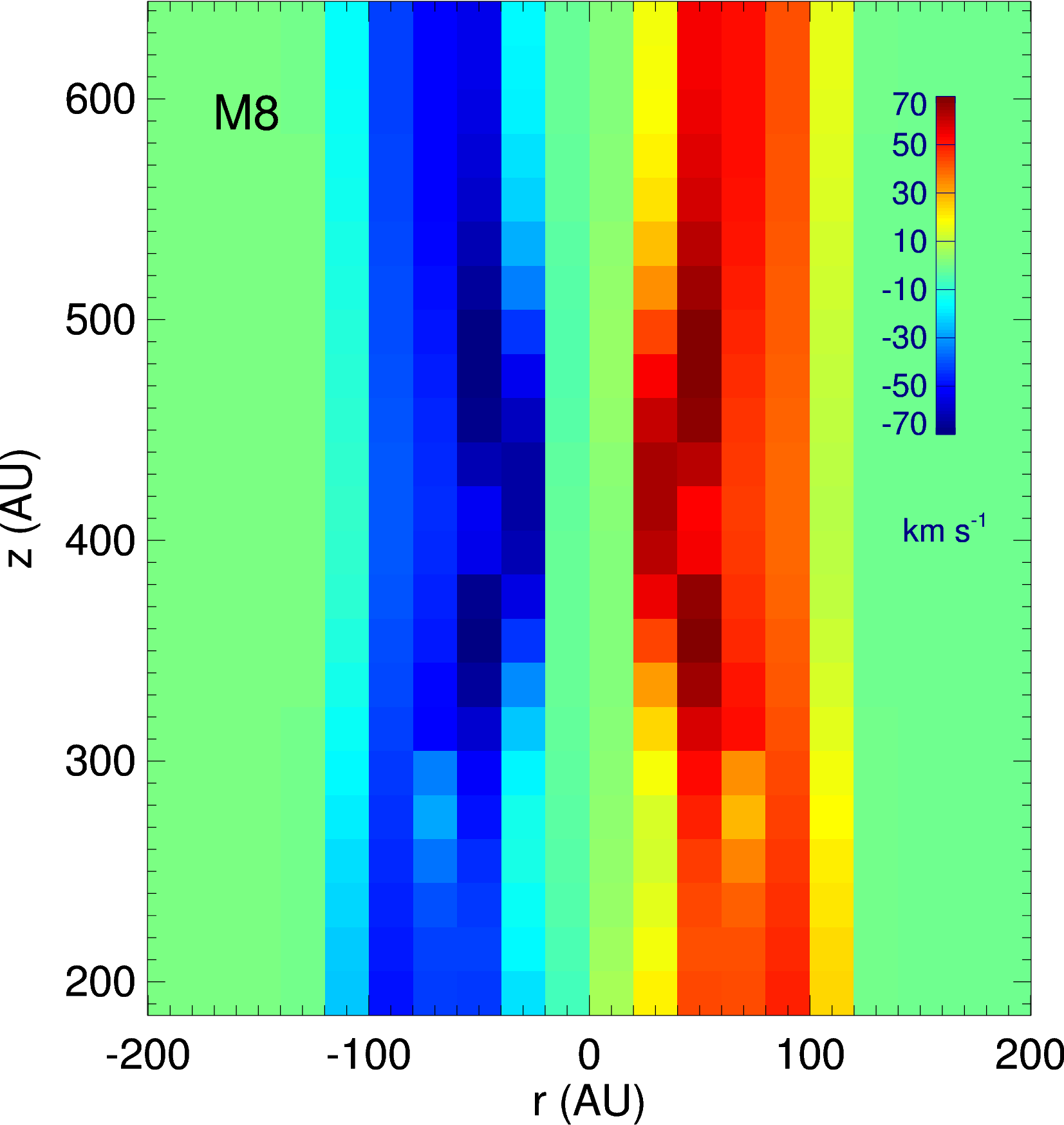}
      \caption{Density-weighted rotational velocity along the line of sight 
      with macropixel resolution of 20 AU after $\approx 50$ years of 
      evolution. Different models are compared: M6 (the model with the lowest 
      rotational velocity) left panel, M7 (the model with an intermediate 
      rotational velocity) middle panel, and M8 (the model with the highest 
      rotational velocity) right panel.}
      \label{twist_comp}
   \end{figure*}

   From the models, we also calculate the density-weighted average velocity 
   along the line of sight and compare it with the velocities inferred from 
   the observations \citep{bac02,cof04,cof07,woi05}. We consider the line of 
   sight perpendicular to the jet axis and we degrade the spatial resolution 
   of the maps derived from the simulations from 0.5 AU to 20 AU (typical 
   average resolution achieved in observations). In Figure~\ref{twist_comp},
   we show two-dimensional distributions for the density-weighted velocities 
   along the line of sight for M6, M7 and M8 obtained with spatial resolution 
   of 20 AU. We find similar characteristics in the three cases. The 
   projected velocities are larger at the edge of the jet where strong 
   $B_{\varphi}$ and $B_{\mathrm{r}}$ components are generated 
   collimating the jet. The velocity reaches its maximum $v_{\mathrm{max}}$ 
   in proximity of the first shock diamond where 
   $v_{\mathrm{max}} \approx 60$~km s$^{-1}$ for model M6, and 
   $v_{\mathrm{max}} \approx 70$~km s$^{-1}$ for models M7 and M8. The values 
   derived in the pre-shock region are quite lower: maximum values for M6, M7 
   and M8 are $\sim 30$~km s$^{-1}$, $\sim 40$~km s$^{-1}$ and 
   $\sim 50$~km s$^{-1}$ respectively. The velocities are significantly 
   lower close to the jet axis where $v_{\mathrm{max}} <10$~km s$^{-1}$. All 
   these values are compatible with those inferred from the observations by 
   \cite{bac02}.

\subsection{Role of thermal conduction and radiative losses}

   We perform some additional simulations for the reference case (see 
   Sect.~\ref{sec:ref}) in order to investigate more deeply the role of 
   thermal conduction and radiative losses in the model. We perform three 
   extra simulations: M0-id, the model M0 for the ideal case (without 
   thermal conduction and cooling); M0-rd, the model M0 including only the 
   radiative cooling; and M0-th, the model M0 including only the thermal 
   conduction. We show the main physical parameters resulting from the 
   different simulations performed in Table~\ref{results_m0}. The jet 
   morphology and the shock position are similar in all the cases, although 
   other shock properties show relevant differences.

   \begin{table*}
      \caption[]{Summary of the main physical parameters resulting from the
      different simulations performed for the reference case, M0 (see 
      Table~\ref{parameters}): shock position (shock starting position from 
      the beginning of the domain), $d_{\mathrm{s}}$, shock maximum density, 
      $n_{\mathrm{s,max}}$, shock maximum temperature, $T_{\mathrm{s,max}}$, 
      shock temperature (calculated as the density-weighted average 
      temperature), $T_{\mathrm{s}}$, shock mass, $m_{\mathrm{s}}$, and shock 
      X-ray luminosity, $L_{\mathrm{X}}$. For the calculation of the shock 
      temperature and mass, $T_{\mathrm{s}}$ and $m_{\mathrm{s}}$ 
      respectively, we only consider the cells with $T\geq 10^6$ in order to 
      describe the shock X-ray emission contribution. M0-id is the model M0 
      for the ideal case (without thermal conduction and cooling), M0-rd is 
      the model M0 including only the radiative cooling, and M0-th is the 
      model M0 including only the thermal conduction.}
      \label{results_m0}
      \centering
      \begin{tabular}{p{1cm} cccccc}
      \hline\hline
      Model  &  $d_\mathrm{s}$ (AU)  &  
      $n_{\mathrm{s,max}}$ ($10^{4}$ cm$^{-3}$)  &  
      $T_{\mathrm{s,max}}$ (MK)  &  $T_{\mathrm{s}}$ (MK)  &  
      $m_{\mathrm{s}}$ ($10^{-9} M_{\odot}$)  &  
      $L_{\mathrm{X}}$ ($10^{28}$ erg s$^{-1}$) \\
      \hline
      M0  &  $340$  &  $1.7$  &  $5.8$  &  $2.1$  &  $1.7$  &  $8.8$ \\
      \hline
      M0-id  &  $340$  &  $0.8$  &  $4.9$  &  $1.5$  &  $3.4$  &  $35.3$ \\
      M0-rd  &  $340$  &  $1.7$  &  $5.0$  &  $2.1$  &  $1.8$  &  $9.1$ \\
      M0-th  &  $340$  &  $0.8$  &  $6.8$  &  $1.5$  &  $3.3$  &  $34.8$ \\
      \hline
      \end{tabular}
   \end{table*}
   
\begin{figure}
      \resizebox{\hsize}{!}{\includegraphics*{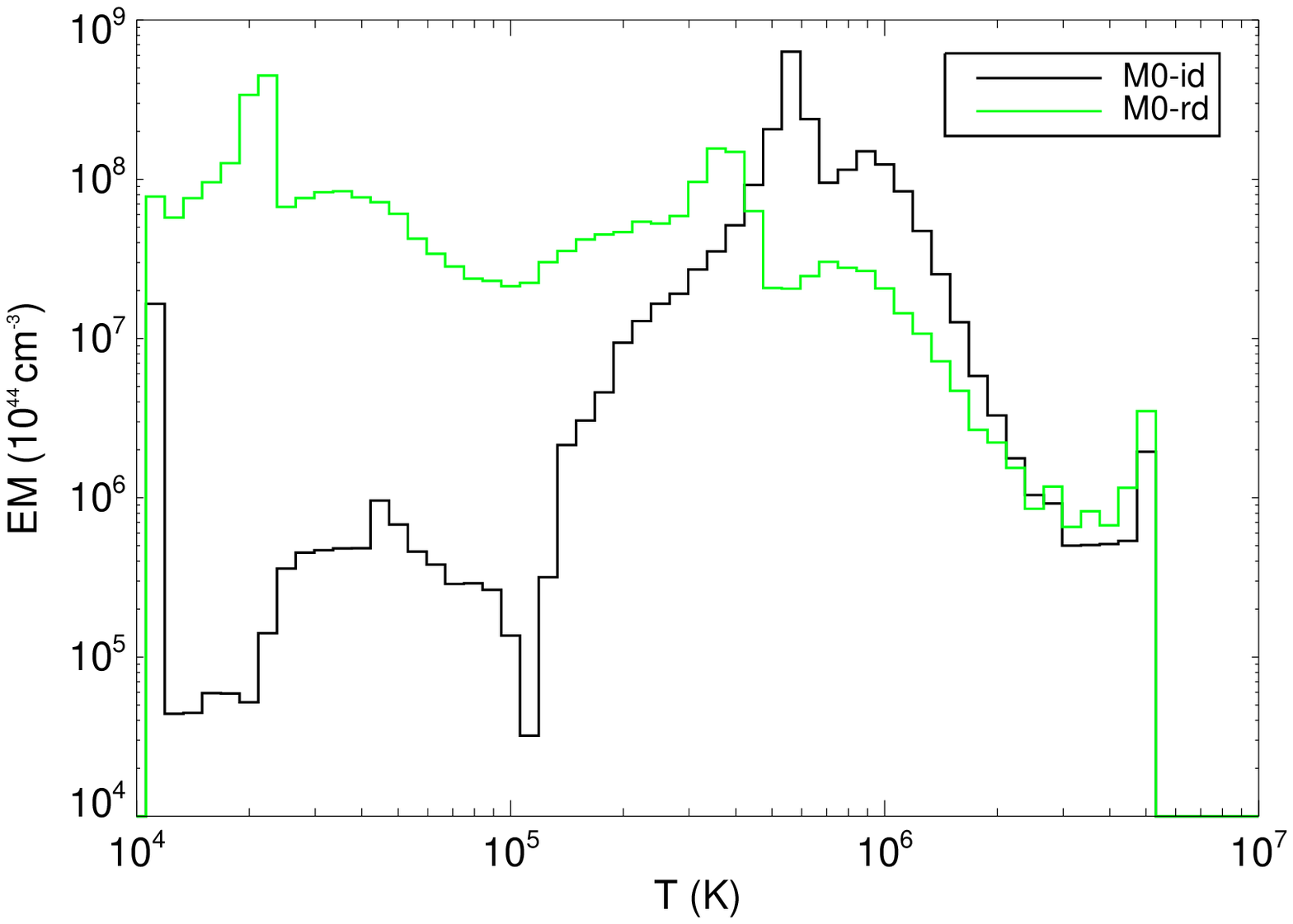}}
      \resizebox{\hsize}{!}{\includegraphics*{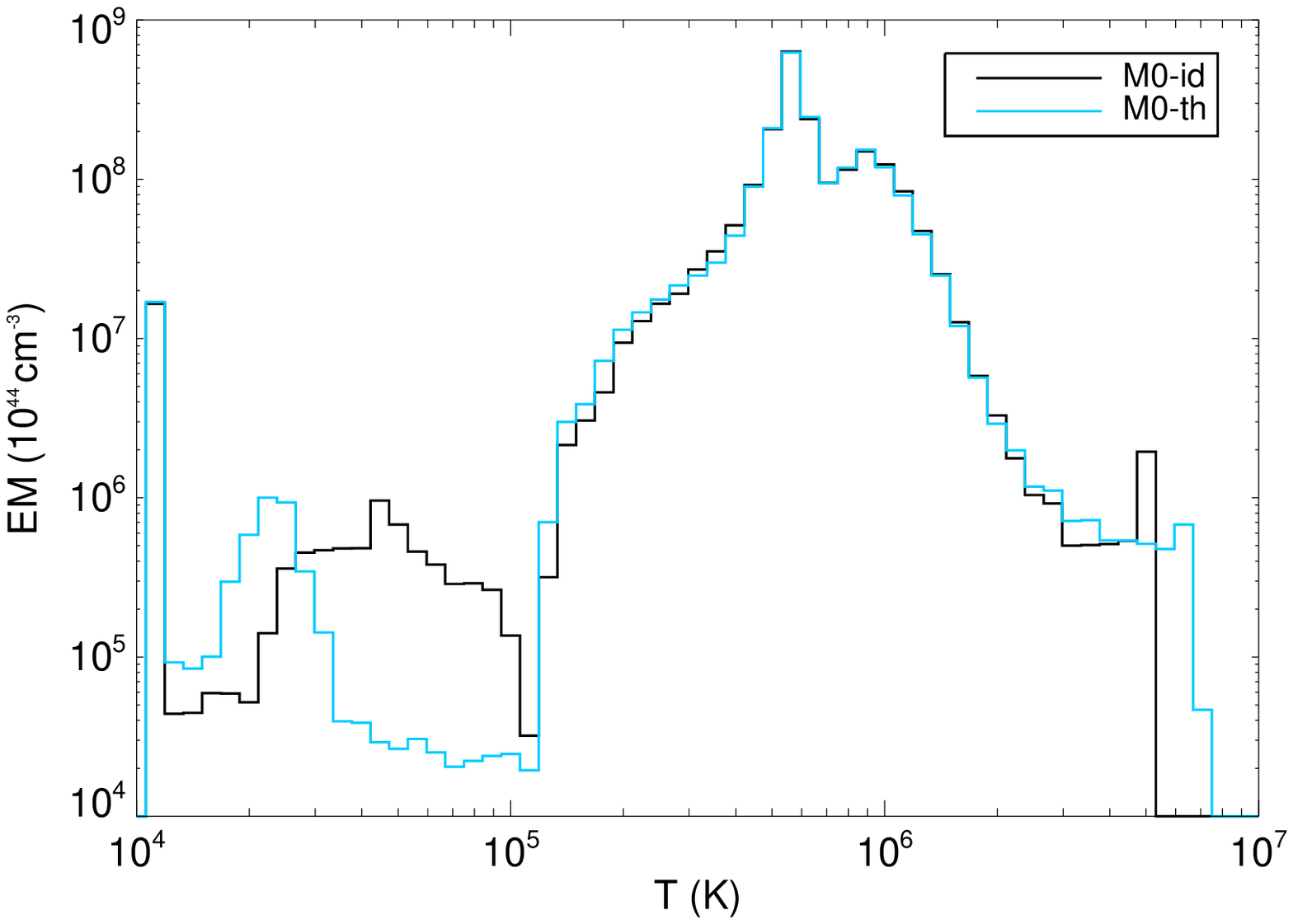}}
      \resizebox{\hsize}{!}{\includegraphics*{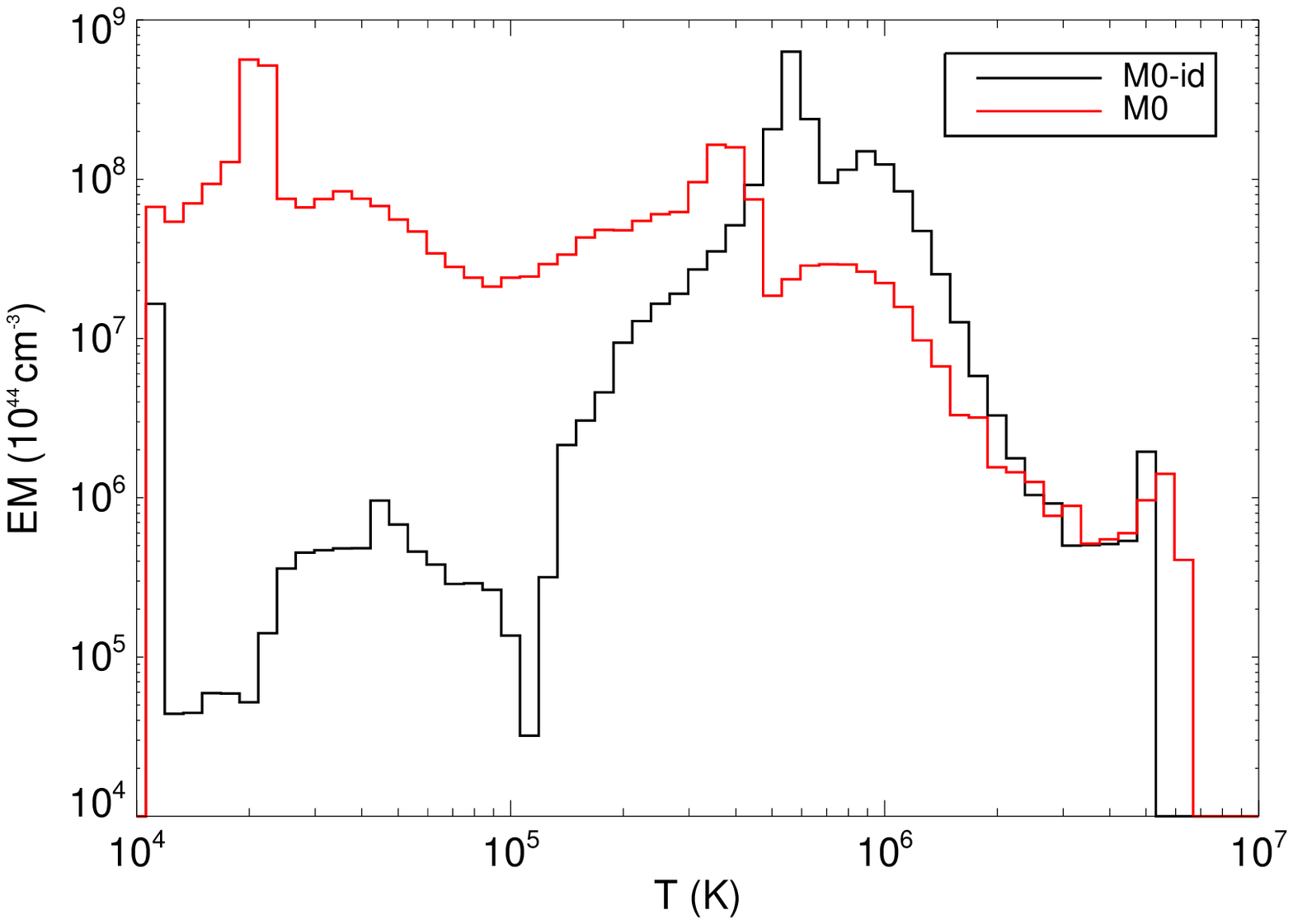}}
      \caption{Emission measure, $EM$, distribution as a function of the 
      temperature, $T$, after $\approx 50$ years of evolution. Different 
      models are compared: M0-rd in green (upper panel), M0-th in blue 
      (middle panel), and M0 in red (lower panel). The $EM(T)$ for the 
      model M0-id is represented by black line in all panels.}
      \label{m0_emtemp_comp}
   \end{figure}

\begin{figure*}
      \includegraphics*[width=0.33\textwidth]{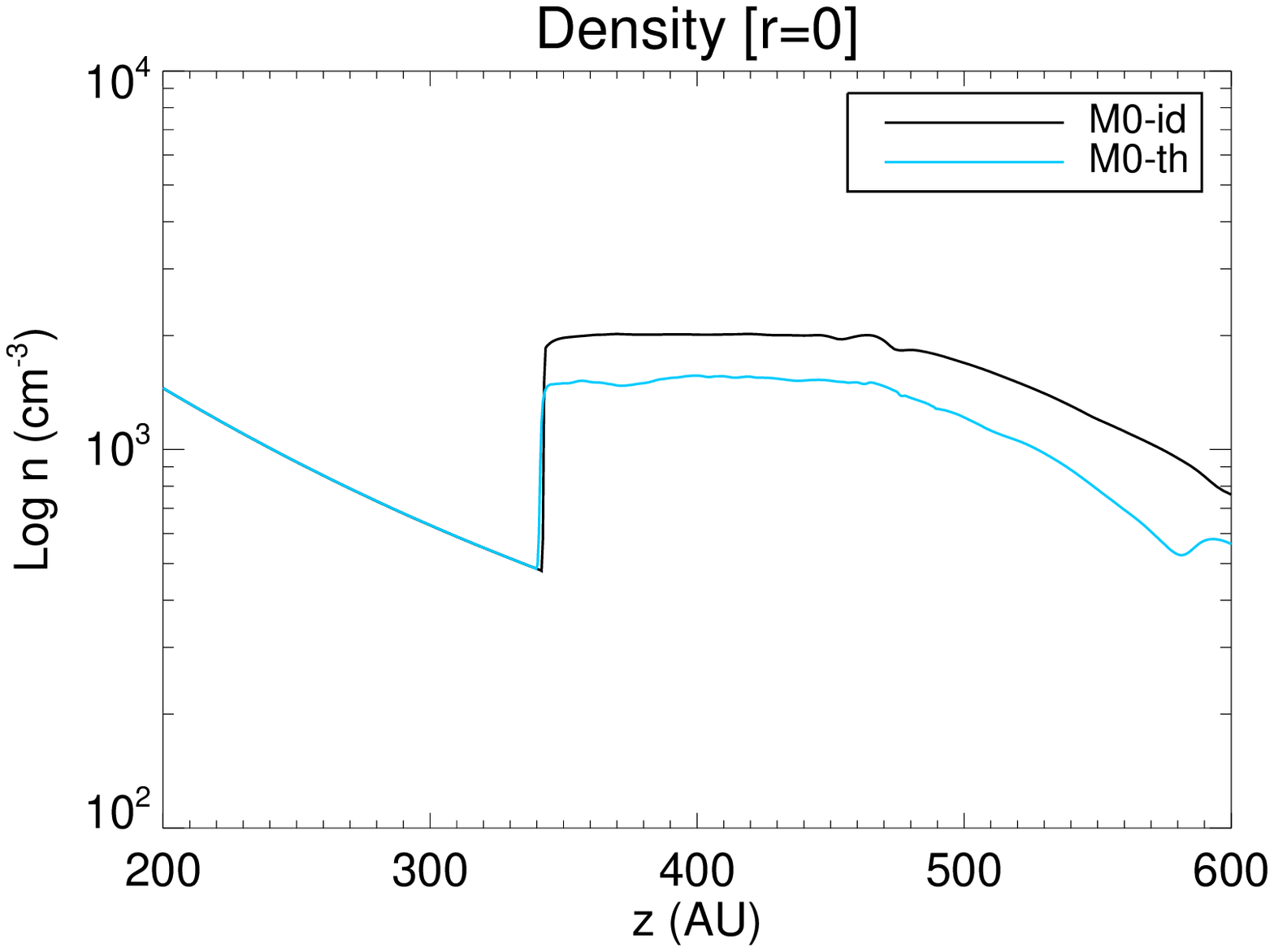}
      \includegraphics*[width=0.33\textwidth]{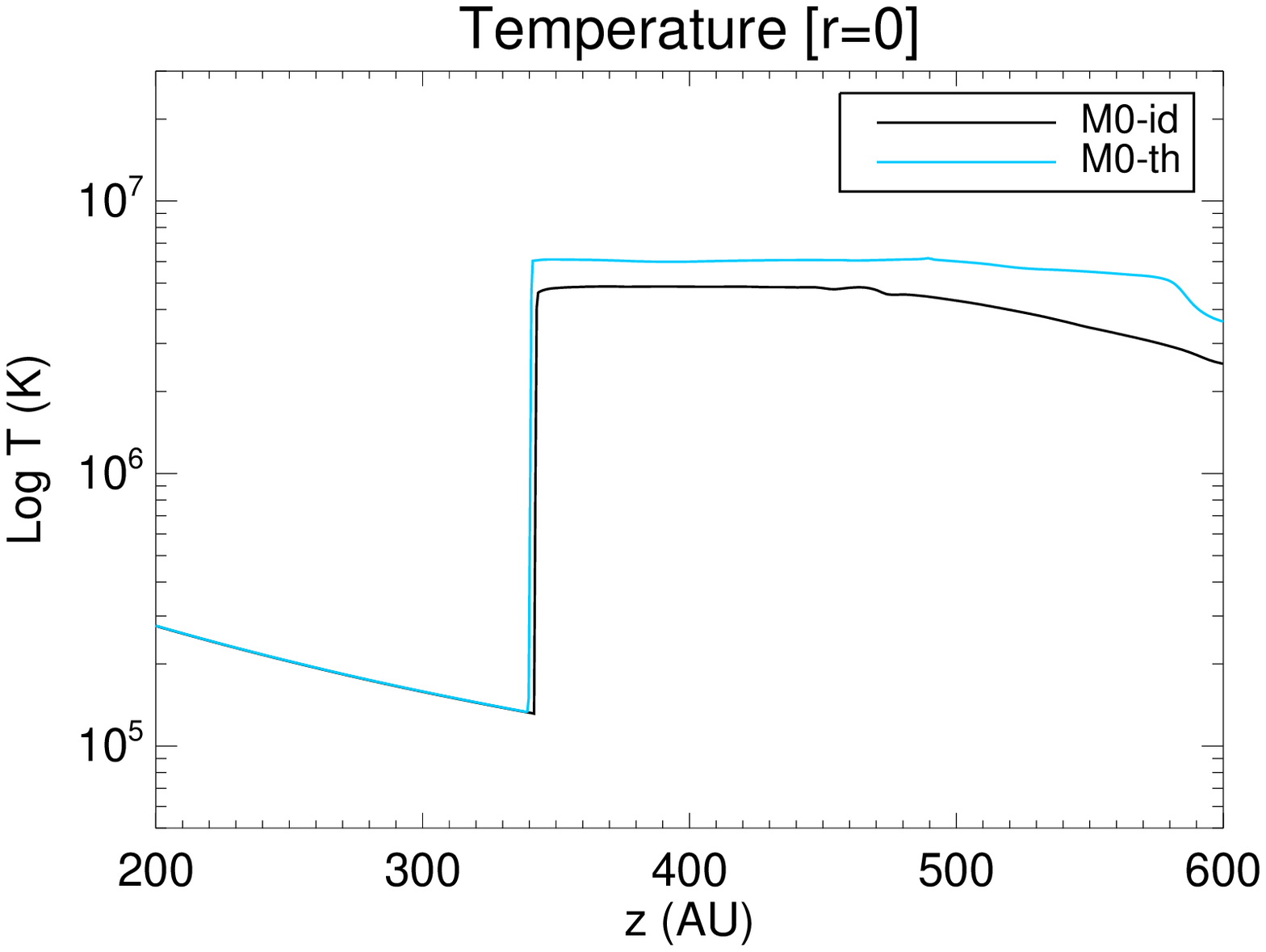}
      \includegraphics*[width=0.33\textwidth]{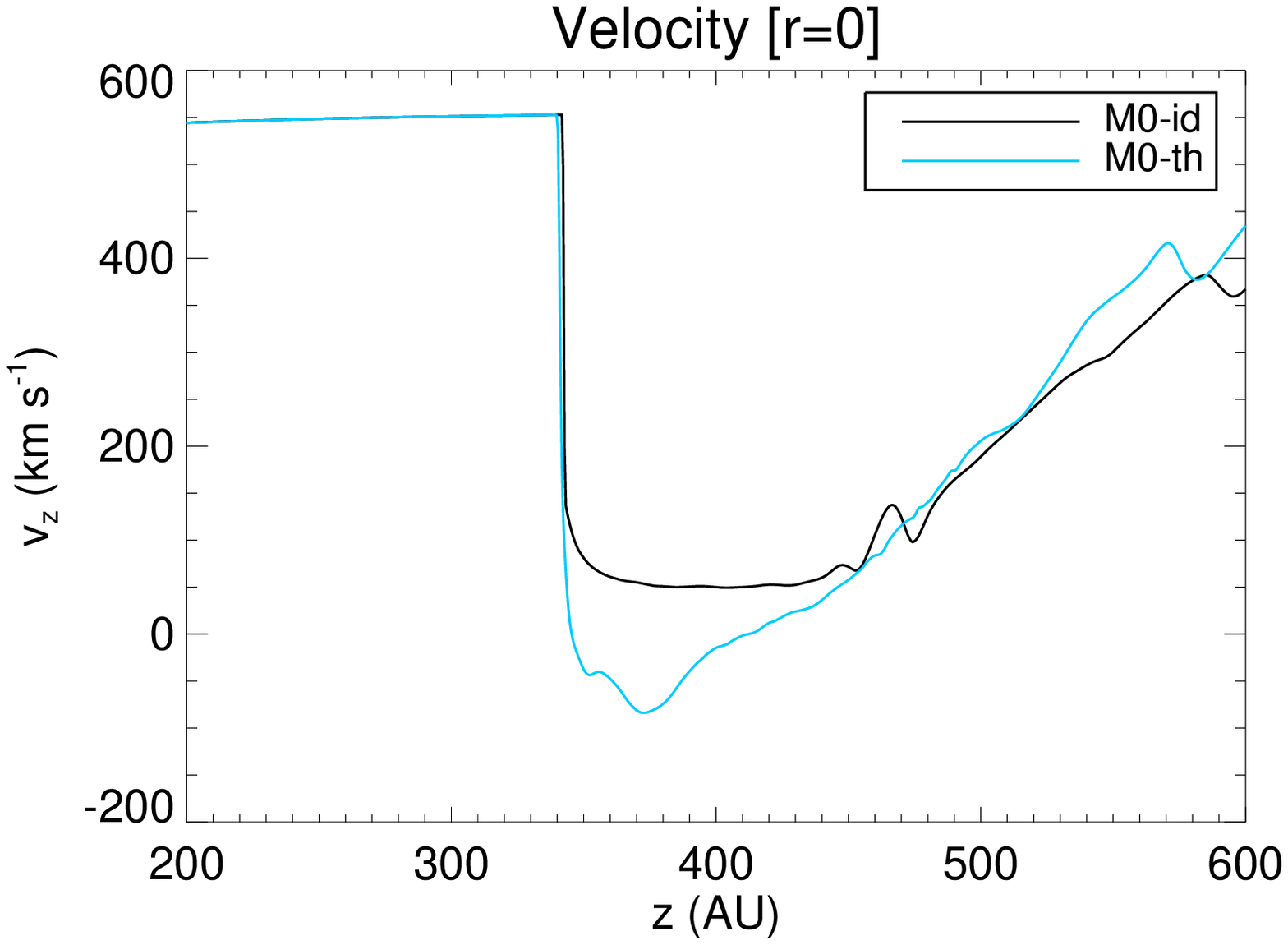}
      \caption{Density (left panel), temperature (middle panel) and velocity 
      (right panel) profiles at $r = 0$ and $t \approx 50$ yr. We compare the 
      ideal MHD model (M0-id) in black with the model including the thermal 
      conduction (M0-th) in blue.}
      \label{prof_comp}
   \end{figure*}

   In Figure~\ref{m0_emtemp_comp} we compare the emission measure 
   distribution vs. temperature, $EM(T)$, for the different models. The 
   upper panel in the figure highlights the effect of radiative losses on 
   the structure of the shock diamond by comparing models M0-id and M0-rd: 
   the effects of cooling determines a decrease of $EM$ at $T> 10^6$~K and an 
   increase of $EM$ at lower temperatures. This is due to the gradual cooling 
   of the plasma in the shock diamond. The middle panel in 
   Fig.~\ref{m0_emtemp_comp} shows the effect of the thermal conduction by 
   comparing models M0-id and M0-th. In this case we note that the thermal 
   conduction has a rather marginal effect in shaping the $EM(T)$ 
   distribution, slightly increasing the $EM$ of plasma with $T > 10^5$~K and 
   decreasing the $EM$ at $T< 10^5$~K. In Figure~\ref{prof_comp} we compare 
   the profiles for density, temperature and velocity for models M0-id and 
   M0-th. The thermal conduction leads to higher values of temperature in the 
   post-shock region, although the pre-shock temperature is the same in both 
   cases (see center panel in Fig.~\ref{prof_comp}). This difference is also 
   reflected in the velocity profiles which show lower values for model M0-th 
   at the shock, even with negative values (see right panel in 
   Fig.~\ref{prof_comp}), forming a vortex shape velocity field. By comparing 
   models including the radiative losses, namely runs M0 and M0-rd (see upper 
   and lower panels in Fig.~\ref{m0_emtemp_comp}), we find that the 
   corresponding $EM(T)$ distributions show similar shapes, thus 
   demonstrating that the structure of the jet is largely dominated by the 
   radiative cooling. The main differences between the two appear for high 
   temperatures ($> 10^6$~K), corresponding to the shock diamond area where 
   the thermal conduction is more efficient and has some effect.

\begin{figure*}
      \includegraphics*[width=0.33\textwidth]{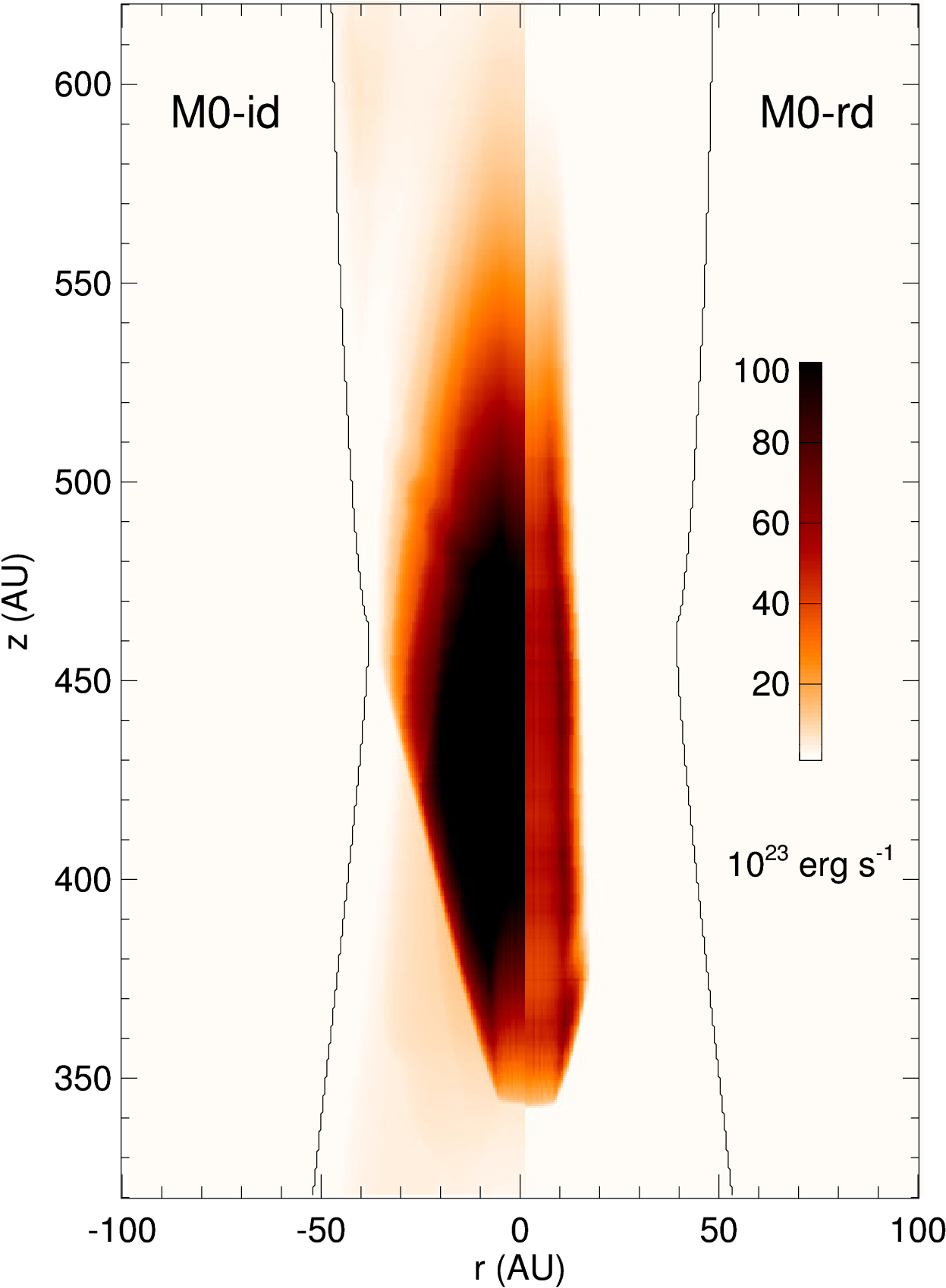}
      \includegraphics*[width=0.33\textwidth]{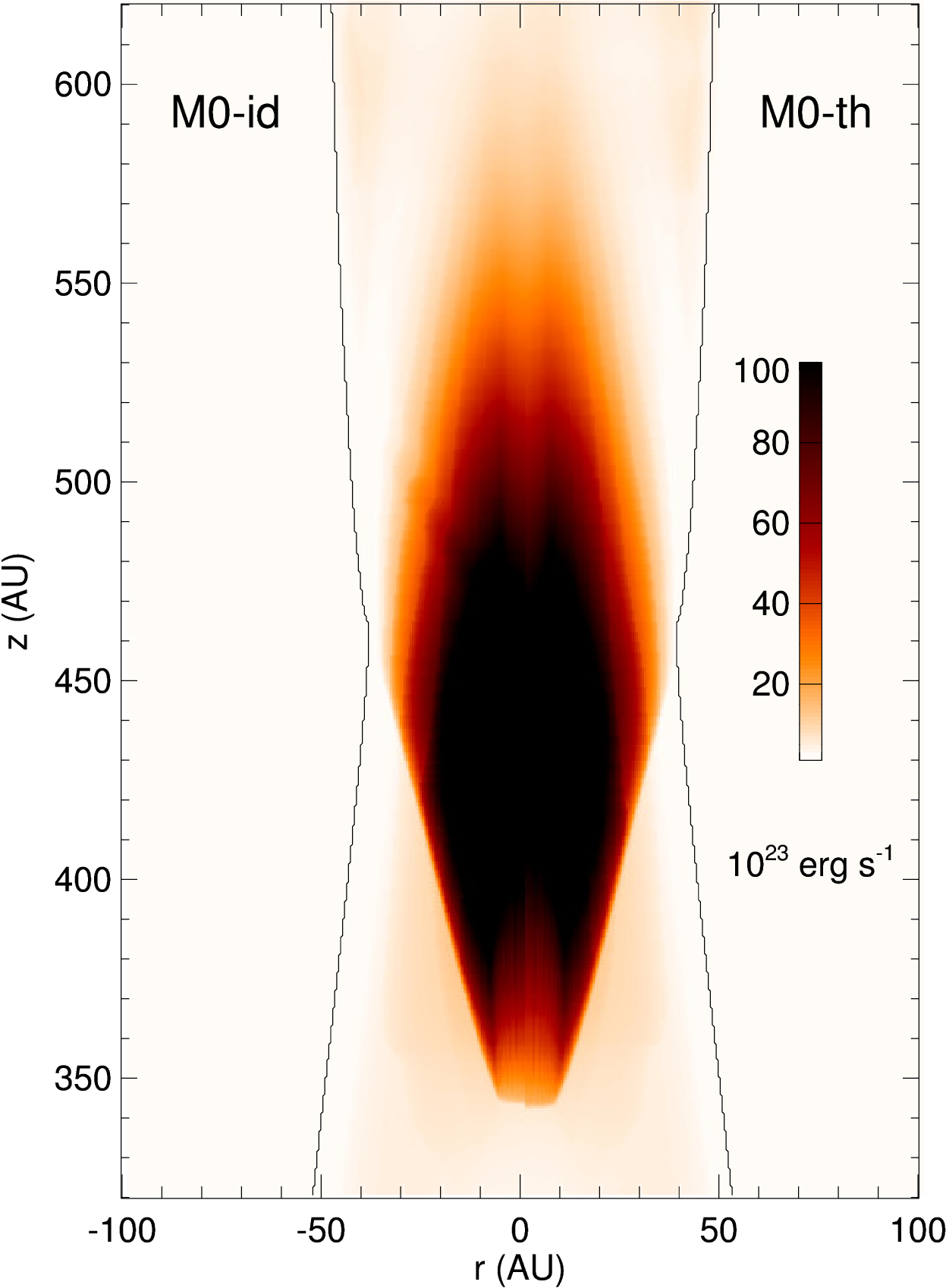}
      \includegraphics*[width=0.33\textwidth]{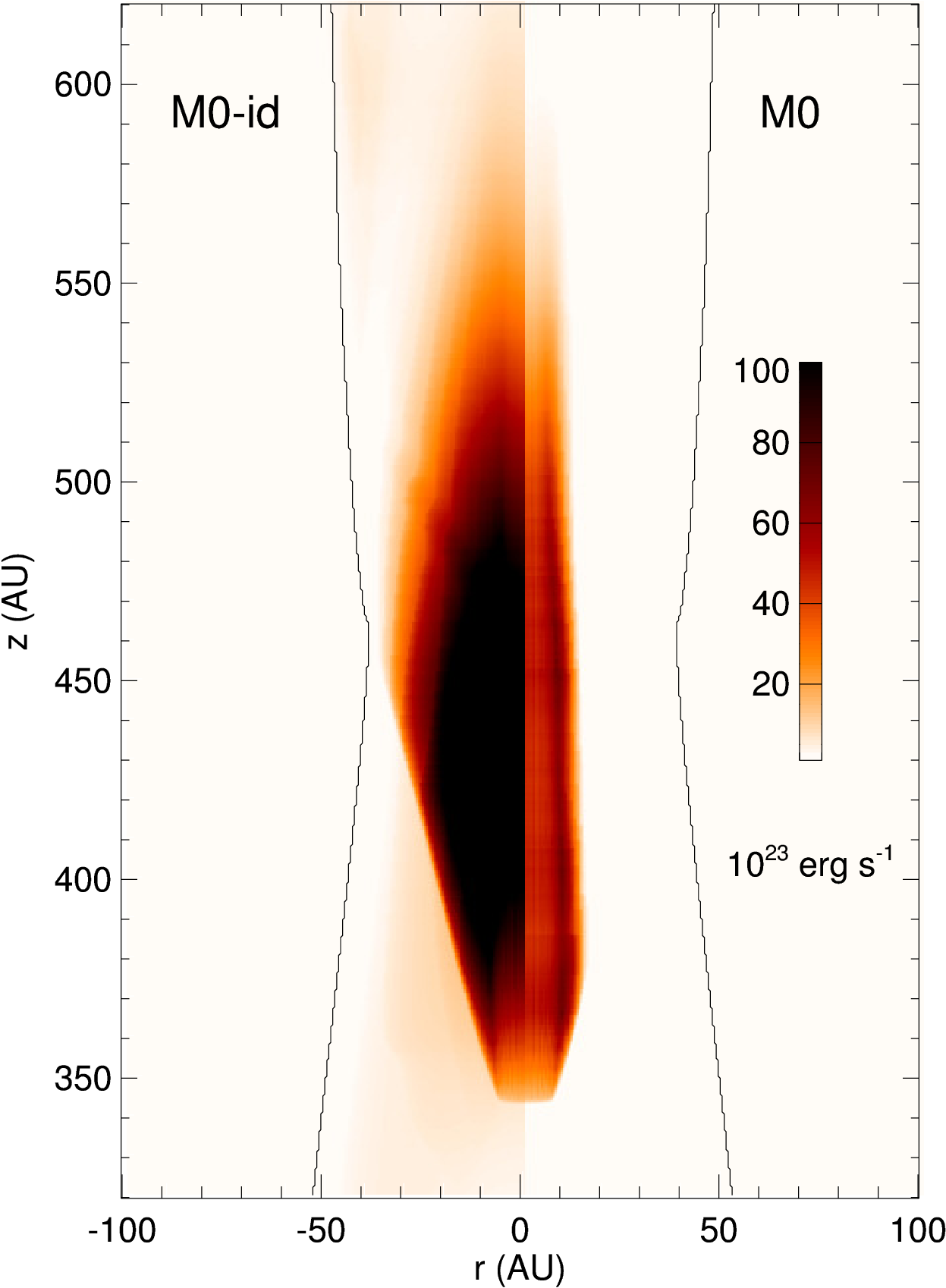}
      \caption{Spatial distribution of the X-ray emission after $\approx 50$ 
      years of evolution. Different models are compared: M0-rd (the model 
      with radiative cooling) right half-panel on the left, M0-th (the model 
      with thermal conduction) right half-panel in the middle, and M0 (the 
      reference case) right half-panel on the right. The X-ray distribution 
      for the model M0-id is reported in each left half-panel.}
      \label{m0_emlum_comp}
   \end{figure*}

   For the different models, we obtain slightly different values for the 
   temperature and density in the shock diamond (see Table~\ref{results_m0}), 
   and we find that the corresponding synthetic luminosities are lower 
   ($\sim 10^{29}$ erg s$^{-1}$) for models with radiative cooling 
   (M0 and M0-rd). The models not considering radiative cooling 
   (M0-id and M0-th) show more extended X-ray sources (see 
   Figure~\ref{m0_emlum_comp}) and higher luminosities 
   ($\sim 3\cdot 10^{29}$ erg s$^{-1}$).

\section{Discussion and conclusions}

   The analysis of the observations of HH154 in three different epochs with 
   Chandra \citep{bal03,fav06,sch11,bon11} revealed a faint and elongated 
   X-ray source at some 150 AU from the protostar that appeared to be 
   quasi-stationary over a time base of $\sim 8$ yr without appreciable 
   proper motion and variability of X-ray luminosity and temperature. 
   Also, the best-studied bright, central X-ray jet of DG Tau seemed to be
   stationary on timescales of several years on spatial scales of about 30 
   AU from the central star \citep{sch08}. One of the viable models proposed 
   by \citet{bon11} on the base of hydrodynamic modelling, explained the 
   observations of HH154 describing a shock diamond formed at the opening 
   of a nozzle and producing a X-ray stationary source. They propose a 
   magnetic nozzle as the origin of the shock and they derive a magnetic 
   field strength $B \approx 5$ mG in the magnetic nozzle at the base of the 
   jet.

   Following the above line of research, we proposed here a new MHD model 
   which describes the propagation of a jet through a magnetic nozzle which 
   rams with a supersonic speed into an initially isothermal and homogeneous 
   magnetized medium. Our MHD model takes into account, for the first time, 
   the relevant physical effects, including the radiative losses from 
   optically thin plasma and the magnetic field oriented thermal conduction.

   We investigated how the magnetic nozzle contributes to the jet collimation 
   and, possibly, to the formation of a shock diamond at the exit of the 
   nozzle. To this end, we performed an extensive exploration of the 
   parameter space that describes the model. These results allowed us to 
   study and diagnose the properties of protostellar jets over a broad range 
   of physical conditions and to determine the physical properties of the 
   shocked plasma. The different parameters considered are shown in
   Table~\ref{parameters}.

   We found that a minimum magnetic field ($\sim1-2$ mG) is necessary to 
   collimate the plasma and form a train of shock diamonds. We selected a 
   magnetic field strength of 5 mG as reference case. For lower values of 
   magnetic field strength, we found that the shock forms at larger distances 
   from the driving star and the beam radius is wider than those usually 
   observed. We found that the stronger is the magnetic field and the lower 
   is the flow velocity, the closer forms the shock to the base of the jet. 
   The summary of results shown in Table~\ref{results} could be very useful 
   in some cases to constrain part of the main physical parameters of 
   protostellar jets from those already known.

   We derived the physical parameters of a protostellar jet that can give 
   rise to stationary X-ray sources at the base of the jet consistent with 
   observations of HH objects. We found that, in most of the cases explored, 
   quasi-stationary X-ray emission originates from the first shock diamond 
   close to the base of the jet. We obtained shock temperatures of 
   $\sim 1-5 \cdot 10^6$~K, in excellent agreement with the X-ray results of 
   \citet{fav02} and \citet{bal03}. X-ray emission from several HH objects 
   was detected with both the XMM-Newton and Chandra satellites: HH2 in Orion 
   \citep{pra01}, HH154 in Taurus \citep{fav02,bal03}, HH168 in Cepheus A 
   \citep{pra05}, and HH80 in Sagittarius \citep{lop13}. 
   They showed luminosities of 
   $L_\mathrm{X} \approx 10^{29}-10^{30}$ erg s$^{-1}$. Other cases 
   well studied, as Taurus jets of L1551 IRS-5, DG Tau, and RY Tau, showed 
   luminous ($L_\mathrm{X} \approx 10^{28}-10^{29}$ erg s$^{-1}$) X-ray 
   sources at distances corresponding to 30-140 AU from the driving star
   \citep{fav02,bal03,gud08,sch11,ski11}. The parameters used in our model 
   and the luminosity values synthesized from the model results are in good 
   agreement with those observed: our reference case predicts a luminosity 
   $L_\mathrm{X} \approx 10^{29}$ erg s$^{-1}$ which is in good agreement with 
   observed values. The cases with higher observed luminosities, as HH154, 
   are in good agreement with models M9-M11 (with lower velocities), and 
   M6-M8 (with magnetic field twisting). We obtain the highest luminosity 
   in the model M8, the model with the highest jet rotational velocity.

   In additon, we investigated the effect of jet rotation on the structure 
   of shock diamonds. Several detections of gradients in the radial velocity 
   profile across jets from T Tauri stars were reported 
   \citep{bac02,cof04,cof07,woi05}. These velocity shifts might be 
   interpreted as signatures of jet rotation about its symmetry axis. For 
   example, \citet{bac02} derived toroidal velocities of the emitting regions 
   between 6 and 15 km s$^{-1}$, depending on position. \citet{woi05} gave 
   higher toroidal velocities in the range 5-30 km s$^{-1}$. They interpreted 
   these velocity asymmetries as rotation signatures in the region where the 
   jet has been collimated but has not yet manifestly interacted with the 
   environment. Considering our model, this region corresponds to the bottom 
   part of the velocity maps in Fig.~\ref{twist_comp} with values 
   $5-30$ km s$^{-1}$ for model M6, $6-40$ km s$^{-1}$ for model M7 and 
   $7-50$ km s$^{-1}$ for model M8.
   Alternative interpretations include asymmetric shocking and/or jet 
   precession \citep[e.g.,][]{sok05,cer06,cor09}.

   We derived the angular momentum loss rate at the base of the jet for the 
   three models, M6, M7 and M8 as 
   $\dot J_{\mathrm{j},\omega} = 
   \int \rho_{\mathrm{j}} v_{\mathrm{j}} v_\varphi r \, \mathrm{d}A$, 
   where $\rho_{\mathrm{j}}$ and $v_{\mathrm{j}}$ are the mass density and 
   jet velocity, respectively, $v_\varphi$ is the rotational velocity, $r$ is 
   the radius and $\mathrm{d}A$ is the cross sectional area of the incoming 
   jet plasma. The values obtained range between 
   $1.63 \cdot 10^{-5}$ $M_{\odot}$ yr$^{-1}$ AU km s$^{-1}$ for model M6 
   and $3.25 \cdot 10^{-5}$ $M_{\odot}$ yr$^{-1}$ AU km s$^{-1}$ for model 
   M8. These values refer to the flux carried away by the jet and 
   they are in good agreement with that estimated by \citet{bac02}, namely, 
   $3.8\cdot 10^{-5}$ $M_{\odot}$ yr$^{-1}$ AU km s$^{-1}$. Thus, our model 
   could be a useful tool for the investigation of the still debated rotation 
   of the jets.

   We also explored the role of thermal conduction and radiative losses in 
   determining the structure of the shock diamonds by performing some extra 
   simulations calculated with these physical processes turned ``on'' or 
   ``off''. We found that the radiative losses dominate the evolution of the
   shocked plasma in the diamonds. The main effect is to decrease the $EM$ of 
   plasma with temperature larger than $10^6$~K, thus reducing its X-ray 
   luminosity. The thermal conduction plays a minor role slightly contrasting 
   the cooling of hot plasma due to radiative losses.

   The comparison between our model results and the observational findings 
   showed that the model reproduces most of the physical properties observed 
   in the X-ray emission of protostellar jets (temperature, emission measure, 
   X-ray luminosity, etc.).  Thus we showed the feasibility of the physical 
   principle on which our model is based: a supersonic protostellar jet leads 
   to X-ray emission from a stationary shock diamond, formed after the 
   jet is collimated by the magnetic field, consistent with the 
   observations of several HH objects.  We conclude that our model provides a 
   simple and natural explanation for the origin of stationary X-ray sources 
   at the base of protostellar jets. Therefore, the comparison of our MHD 
   model results with the X-ray observations could provide a fundamental tool 
   to investigate the role of the magnetic field on the protostellar jet 
   dynamics and X-ray emission.


\begin{acknowledgements}
   S.U. acknowledges the hospitality of the INAF Osservatorio Astronomico di 
   Palermo, where part of the present work was carried out using the HPC 
   facility (SCAN).
   PLUTO is developed at the Turin Astronomical Observatory in collaboration 
   with the Department of General Physics of the Turin University.
   Part of the calculations of this work were performed in the high capacity 
   cluster for physics, funded in part by UCM and in part with Feder funds. 
   This is a contribution to the Campus of International Excellence of 
   Moncloa, CEI Moncloa.
   This work was supported by BES-2012-061750 grant from the Spanish 
   Government under research project AYA2011-29754-C03-01. 
   Financial support from INAF, under PRIN2013 programme `Disks, jets and 
   the dawn of planets' is also acknowledged.
   Finally, we thank the referee for useful comments and suggestions.
\end{acknowledgements}


\bibliographystyle{aa} 
\bibliography{biblio.bib} 

\end{document}